\pdfoutput=1
\PassOptionsToPackage{table,xcdraw}{xcolor}

\documentclass[sigconf]{acmart}
\renewcommand\footnotetextcopyrightpermission[1]{} 
\usepackage{natbib}
\usepackage{xurl}   
\usepackage{url}
\usepackage{graphicx}

\AtBeginDocument{%
  \providecommand\BibTeX{{%
    \normalfont B\kern-0.5em{\scshape i\kern-0.25em b}\kern-0.8em\TeX}}}

\usepackage{float}
\usepackage{algpseudocode}
\usepackage{algorithm}

\copyrightyear{2023}
\begin{document}

\title{Image Pre-processing for Improving Deep Learning Classification of Retinopathy of Prematurity}

\author{Sajid Rahim}
\affiliation{%
   \institution{McMaster University}
   \city{Hamilton}
   \state{Ontario}
   \country{Canada}
}
\author{Dr Kourosh Sabri}
\affiliation{%
   \institution{McMaster Hospital}
   \city{Hamilton}
   \state{Ontario}
   \country{Canada}
}
\author{Dr Anna Ells}
\author{}
\affiliation{%
  \institution{University of Calgary}
  \city{Calgary}
  \state{Alberta}
  \country{Canada}
}
\author{Dr Alan Wassyng }
\affiliation{%
   \institution{McMaster University}
   \city{Hamilton}
   \state{Ontario}
   \country{Canada}
}
\author{Dr Mark Lawford}
\author{}
\affiliation{%
  \institution{McMaster University}
  \city{Hamilton}
  \state{Ontario}
  \country{Canada}
}
\author{Dr Wenbo He}
\author{}
\affiliation{%
  \institution{McMaster University}
  \city{Hamilton}
  \state{Ontario}
  \country{Canada}
}
\author{Dr Linyang Chu}
\author{}
\affiliation{%
  \institution{McMaster University}
  \city{Hamilton}
  \state{Ontario}
  \country{Canada}
}


\renewcommand{\shortauthors}{Author 1 et al.}

\begin{abstract}
Retinopathy of Prematurity (ROP) can affect babies born prematurely. It is a potentially blinding eye disorder because of damage to the eye’s retina. Screening of ROP is essential for early detection and treatment. This is a laborious and manual process which requires a trained physician performing a dilated ophthalmology examination. This procedure can be subjective resulting in lower diagnosis success for clinically significant disease. Automated diagnostic methods using deep learning can help ophthalmologists increase diagnosis accuracy by utilising patient's retinal images also known as fundus images that can be digitally captured with Retcam\cite{retcam2023} or smartphones. In pediatrics ophthalmology, it is difficult to capture Retcam images from a premature infant. Captured Retcam images are challenged with poor quality due to many factors that mandates image pre-processing prior to usage in automated diagnosis. Most research groups have employed traditional image processing to improve Retcam image quality for use in deep learning classification for ROP conditions. This is effective to a limited degree. Motivated by the quality of pediatric Retcam images, this paper proposes two improved novel restoration image pre-processing methods as well as exploring a novel way using segmentation to erode blood vessels in the images. These combined with traditional methods also further improve ROP features obtained by Retcam images for use with pre-trained transfer learning CNN frameworks to create hybrid models that result in higher diagnosis accuracy. We created a set of Deep learning classifiers for Plus, Stages and Zones. These were trained and validated using the improved pre-processing methods and traditional methods independently. 
Our evaluations showed that these new methods contributed to higher accuracy than traditional image pre-processing when applied to our ROP Retcam datasets. Further, our results were as equal or better than comparative peer results using limited data.
\end{abstract}



\maketitle
\pagestyle{plain}
\thispagestyle{empty}
\fancyfoot{}

\makeatletter
\renewcommand\@formatdoi[1]{\ignorespaces}
\makeatother

\section{Introduction}
\label{sec:introduction}
Retinopathy of Prematurity (ROP) is a disorder of the developing retinal blood vessels in premature infants and is a leading cause of childhood blindness. In full term infants, ROP does not occur as the retinal vasculature is fully developed. In premature infants, the development of the retinal blood vessels, which proceeds peripherally from the optic nerve during gestation, is incomplete. Hence, the extent and possibility of immature development of the retina depends on the degree of prematurity. \cite{fierson2018screening} \cite{section2006screening}.

\begin{figure}[ht!]
    \centering
    \includegraphics[height=2in,width=3.2in]{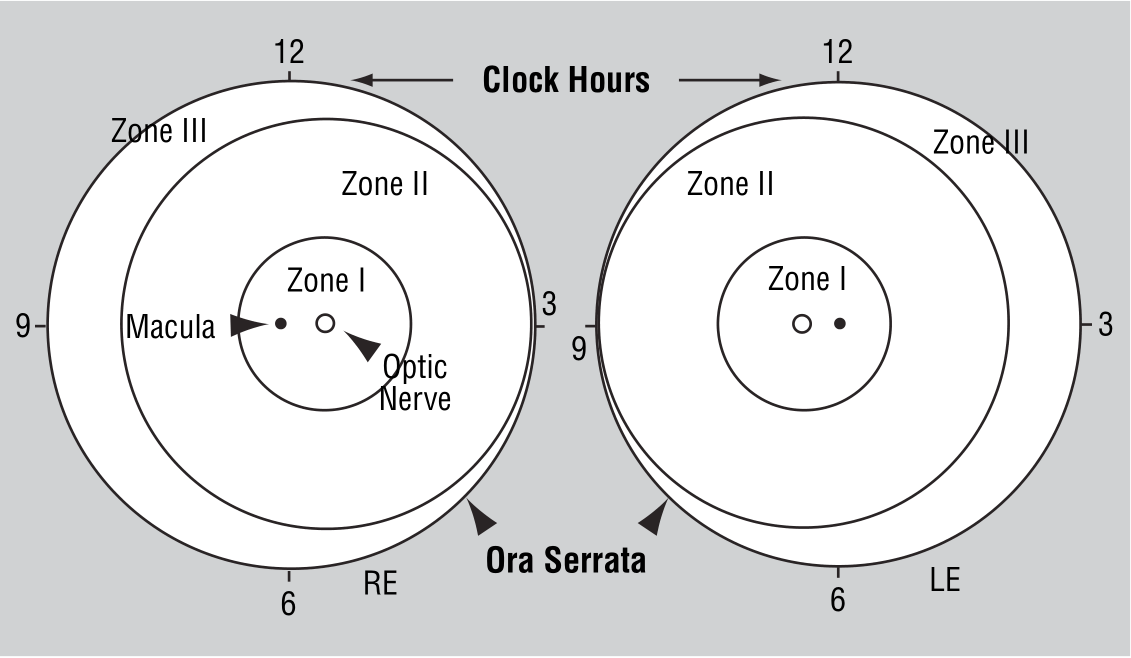}    
    \caption{Retina showing zones/hours for location and extent of ROP \cite{jefferies2016retinopathy}}
    \label{fig:ClassifyROP}
    \Description{ClassifyROP}
\end{figure}

There are several stages or classification of Retinopathy of Prematurity. These are based on the schema of the retina in the left and right eye (Fig ~\ref{fig:ClassifyROP}). They can be summarised as noted by the American Academy of Ophthalmology \cite{stephen2022} as follows: 

\begin{enumerate}
    \item \textbf{Location:} 
    {\renewcommand\labelitemi{}
    \begin{itemize}
        \item \textit{Zone I} – posterior retina within a 60 degree circle centred on optic nerve.
        \item \textit{Zone II} – from posterior circle (Zone I) to nasal ora serrata anteriorly.
        \item \textit{Zone III} – remaining temporal peripheral retina.  The extent is indicated by the number of \textit{clock hours} involved (see Fig ~\ref{fig:ClassifyROP}). 
    \end{itemize}
    }
    \item \textbf{Severity:} 
        {\renewcommand\labelitemi{}
        \begin{itemize} 
            \item \textit{Stage 0} – immature retinal vasculature without pathologic changes.
            \item \textit{Stage 1} - presence of demarcation line between vascularized and non-vascularized retina. 
            \item \textit{Stage 2} - presence of demarcation line having a ridge.
            \item \textit{Stage 3} - a ridge with extra-retinal fibrovascular proliferation.
            \item \textit{Stage 4} - partial retinal detachment.
            \item \textit{Stage 5} - total retinal detachment.
        \end{itemize}
        }
    \item \textbf{Plus disease:}
    {\renewcommand\labelitemi{}
    \begin{itemize}
        \item Vascular dilation and tortuosity of posterior retinal vessels in at least 2 quadrants of the eye.
    \end{itemize}
    }
\end{enumerate}

Clinical diagnosis remains a challenge as it requires direct dilated ophthalmological examination and there is variability between experts in diagnosing various stages of ROP. There are very few highly trained specialists able and willing to manage ROP in most countries. This impacts patients who may be located across a large geographic area such as Canada. This is additionally hampered by the fact that training of ROP specialists is time intensive. Examining patients is itself a challenge given the age of these premature infants. ROP incidents worldwide have been increasing given the improvements of neonatal care and population growth \cite{sabri2016global, quinn2016retinopathy,reid2019artificial}.

Globally, 19 million children suffer from visual impairment of which 1.85 million are likely to have developed some degree of ROP. For those diagnosed with ROP, 11\% develop severe visual impairment, and 7\% develop mild/moderate impairment. ROP incidence rates have been estimated as  9\% in developed countries but higher at 11\% in developing countries \cite{sabri2016global}. 

Early detection is critical so that appropriate follow-ups and treatments can be provided. The goal is to improve patient care by streamlining the screening of patients at risk of ROP. One study reported 55 infants examined for every one treated in the UK \cite{haines2002retinopathy}. Providing improved quality of care by enabling specialists to remotely diagnose patients can be very beneficial, especially in remote locations in Canada and  many parts of the world. An automated computer-based image analysis/diagnosis of ROP using captured retinal examination images adds to the support system of early detection. 

Retcam \cite{retcam2023} is the preferred paediatric ophthalmological camera system. Recently, Smartphone cameras using MilRetCam \cite{miiretcam2023} are also being used. Different capturing systems have different resolution and depth quality that requires standardisation for usage in deep learning especially with intermixed sources. ROP Retcam image quality is a significant challenge. Capturing quality images is extremely challenging with premature babies. Over saturation or illumination of images is a consistent problem. Poor image quality occurs also from degeneration within the captured image that results from Retcam lens to cornea contact air gaps, as well as reflection from the cornea, vitreous, and the retina itself. There is also image variability due to fundus pigmentation that  varies across ethnicity and geographic region. Physiologically, infants have more visible choroid vessels which can challenge classification \cite{reid2019artificial}.  Meeting these challenges is essential for clinicians considering using machine learning techniques in clinical practice for screening, diagnosing and monitoring post treatment. In this work, we demonstrated how improved image pre-processing can overcome ROP Retcam image quality  for use with convolutional neural networks (CNNs) or in clinical use.

Review of relevant literature \ref{sec:ROPCNNs}shows that CNNs are the most effective deep learning methods for this diagnosis/classification task, so throughout this paper we focused only on CNN based methods. When using CNN for ROP classification, researchers primarily used traditional image pre-processing methods to overcome ROP quality challenges. In this paper, we proposed novel pre-processing for improving ROP Retcam image quality. When combined with existing traditional image pre-processing techniques, these merged hybrid techniques improved salient ROP features. We highlighted the resizing problem for ROP Retcam images with existing algorithms available with recommendation for ROP Retcam images for CNN use. We then conclusively demonstrated that improved the quality of ROP Retcam images using the improved novel and hybrid methods led to better diagnosis of ROP disease in terms of quality and consistency when used with a CNN based classifier. These pre-processed images can also be used in a clinical setting to improve practical ROP diagnosis and classification. 

In this research, we focused on the presence of Plus Disease, Stages 0-3, and Zones I-III. Stages 0-3 was chosen to detect the presence, outline and depth of the demarcation line as Stages 1-3 are most critical for patient care. As Stages 4-5 data was unavailable, and therefore not evaluated. We created separate training and validation datasets for Plus disease, Stages 0-3, and Zones. Each pair of ROP Retcam dataset was pre-processed using the same pre-processing method or a combination. Two independent transfer learning based CNN classifiers were used separately to predict Plus disease, Stages of ROP and Zones, independent of each other. Training, and validation was repeated for each different pre-processed image data set using these 2 separate classifiers for Plus Disease, Stages and Zones for ROP. The final validation results weree compared for each pre-processed data set. These results demonstrated that for the same pair of ROP data sets,  with improved novel image pre-processed methods did improve CNN classification results for Plus Disease, Stages and Zones. 

The primary contribution of this work includes: 
\begin{itemize}
    \item Improved two restoration based methods, namely Pixel Colour Amplification and Double Pass Fundus Reflection, and combined them with CLAHE which significantly improved ROP Retcam quality in comparison to traditional methods. 

    \item Demonstrated a novel approach to fundus feature reduction by using a CNN based segmentation to generate a vessel map. The vessel maps were then used to erode the vessels into the background in the original ROP Retcam image.

    \item Demonstrated the impact of the improved methods on ROP Retcam images for CNN by implementing and evaluating images used in two CNN models namely ResNet50, and InceptionResV2 for Plus Disease, Stages (0-3) and Zones (I-III) classifications. These were trained/validated using traditional and our new methods. 
    \item We accurately labelled and sub-categorised our ROP dataset into Plus Disease, Stages, and Zone. This data can be made available to researchers.
\end{itemize}

The paper is structured as follows. Section~\ref{sec:Relatedworks} describes related ROP work using deep learning, and two classes of image pre-processing. Section~\ref{sec:DatasetsMethods} details our system architecture (McROP), dataset used, methods used for image pre-processing, augmentation and resizing, and deep learning networks using transfer learning. Section~\ref{sec:Experiments} describes how our architecture was used in our experimental evaluations. Sections ~\ref{sec:Results}, ~\ref{sec:Discussion} review the results and discusses them in comparison with peer works respectively. Lastly, Section~\ref{sec:ConclusionFutureWorks} presents our conclusion and proposed future work based on findings in our research.
\section{Related Work}
\label{sec:Relatedworks}
We briefly discuss related works for most recent ROP specific approaches using CNNs and review existing Fundus pre-processing methods as well as identify challenges which inspired our contribution.

\subsection{ROP detection based CNNs}
\label{sec:ROPCNNs}

Recent approaches with promising results for automated ROP pathology detection are based on convolutional neural networks (CNN). These use ROP Retcam images as input and do not require manual annotation. These methods for ROP detection are capable of coarse-grained classification, such as discriminating severe from mild ROP. They do not specifically assess disease Stages or Zones. While the literature suggests that severe disease rarely develops without changes in posterior pole vasculature, providing additional outputs of the Zone and Stages could improve the interpretability of the system’s assessment.

Worral et al., \cite{worrall2016automated} were the first to use CNN for ROP. Using approximately 1500 images from 35 patients and 347 exams (2-8 images per eye), they utilised a Bayesian CNN approach for posterior over disease presence, and thereafter used a second CNN trained to return novel feature map visualisations of pathologies. They achieved an accuracy of 91.80\% in terms of identifying eyes with ROP versus healthy eyes, with sensitivity of 82.5\% and specificity of 98.30\% per eye.

Wang et al., \cite{wang2018automated} architected the largest automated ROP detection system named DeepROP using Deep Neural Networks (DNNs). The largest dataset at the time used 20,795 images, 3722 cases, and 1273 patients, with clinical labels added by clinical ophthalmologists. ROP detection was divided into ROP identification and grading tasks using two DNN models, namely Id-Net and Gr-Net. They first determined if ROP was present, and if present, they graded ROP either as Minor (Zone II/III and Stage 1/Stage 2) or Severe (threshold disease, Type 1, Type 2, AP-ROP, Stage 4/5). No finer details are determined such as Plus disease, which Stage of ROP and Zones individually. Standard imaging technology was used, with private datasets and there is no published way to reproduce these findings. Images used were of a single population cohort and pre-processing datasets results were not clearly identified prior to training. Their results were validated against 472 screenings from 404 infants, and resulted in identification of ROP with sensitivity of 84.9\%, specificity of 94.9\% and accuracy of 95.6\%. Minor/Severe classification was achieved with sensitivity of 93\%, specificity of 73.6\% and accuracy of 76.4\%. It outperformed 1 out of 3 experts. 

Redd et al., Brown et al., \cite{brown2018automated} devised a two stage CNN i-ROP-DL which combines prediction probabilities via a linear formula to compute an ROP severity score, which can serve as an objective quantification of disease. A similar idea could provide finer grading of Plus disease. The i-ROP-DL deep learning system was the first to detect specific ROP classifications. 6000 images were used and professionally labelled as normal, pre-Plus, Plus ROP stages. Partial/full detachment images were excluded. Two stage CNNs were used. One was to segment and the other was to classify. The classifier classified only Plus disease, and their solution was limited to handling one type of imaging source, as indicated by the authors . They achieved 91\% accuracy with sensitivity and specificity of 93\%, 94\%, respectively.  For Plus disease, they achieved 100\% accuracy, and a maximum of 94\% for Pre-Plus. This outperformed 6 out of 8 ROP experts. 

Mulay et al., \cite{mulay2019early} focused on detection of the demarcation line using a CNN based model, Mask R-CNN which used generated mask to predict ROP Stage 2 classification based on ridge line presence. Image pre-processing was used to overcome poor image quality. Detection accuracy was 88\% with positive predictive value of 90\%/75\% and with negative predictive value of 97\%/42\% in terms of sensitivity/specificity. This demonstrated that image pre-processing assists deep learning classification for ROP Stages. 

Vinekar et al., \cite{vinekar2021development} used deep learning with 42,000 images from a tele-ROP screening in India, with the goal of detecting the presence of Plus disease. Using two test sets that excluded pre-Plus disease, they were able to achieve 95.7\%/99.6\% and 97.8\%/68.3\% in terms of sensitivity/specificity.

Tong et al., \cite{tong2020automated} architected another novel approach to identifying ROP and its stages using 2-layer CNN DL and predict lesion location in fundus images. Datasets of over 36,000 Retcam images were reviewed and labelled by 13 ophthalmologists into training and validation dataset using 90:10 split. Each image was labelled with a classification (ROP severity) and identification label (ROP clinical stage). Normal when no abnormalities, mild for Stage 1/Stage 2 without Plus disease, semi-urgent for Stage 1/Stage 2 with Plus disease, urgent for Stage 3, Stage 4, and Stage 5 with or without Plus disease. Very basic pre-processing which included adding noise and brightness to images was performed, and the dataset was then augmented for volume and image compressed. Resnet 101 CNN using transfer learning was used for classification and Faster R-CNN was used to identify ROP stage, presence of Plus disease and predict objective boundary of lesions. Very good accuracy was achieved by using transfer learning of 88.30\%, 90\%, 95.70\% and 87\% for Normal, mild, semi-urgent and urgent results with additional accuracy of Stage and Plus disease at 95.70\% and 89.60\%. A repeat without transfer showed poor sensitivity and accuracy. They do show breakdown for Stages as well Plus disease which we can compare our results with. As they were unable to differentiate Stage 0 and Stage 1 due to very subtle demarcation line presence, Stage 0 detection was dropped.

Wu et al., \cite{wu2022development} developed 2 deep learning models to demonstrate that a CNN can be used to predict the occurrence and severity of ROP. Two models were specifically designed to look for presence and severity of ROP using ResNet50. Severity was defined as \textit{mild} if the patient had Type II ROP, Zone II Stage 1, Stage 2 ROP without Plus disease and Zone III Stage 1,2, or 3 ROP. Severity was \textit{severe} if the patient had Stage 4 or 5 ROP, Type I or aggressive posterior ROP.  Results for presence of ROP was 100\%/38\% and severity of ROP was 100\%/47\% in terms of sensitivity/specificity. Here too, image pre-processing was not highlighted as having been used.

Ding et al., \cite{ding2020retinopathy} proposed a hybrid architecture focused on localization of the demarcation line and in parallel feeding a black and white reciprocal image as 2 channel input. The  former provides the input of an area of interest. It is fed into traditional CNN. The paper conducted experiments using 2759 images from a local hospital, that were classified by Stages 1-3 only. Images were pre-processed using contrast enhancing features. The pre-processed image was used for object segmentation by applying Mask R-CNN to highlight the demarcation line at pixel level and generate a bounding box around the area of interest, creating a binary mask for the demarcation line. The 299x299 resized image was then used for training with Transference learning to prevent over-fitting. Inception v3 pre-trained on ImageNet was used for classification. Stage 1, 2 and 3 classifications achieved 77\%/78\%, 62\%/61\%, 62\%/62\% in terms of sensitivity/specificity. Of significance is their ability to detect between Stage 1 and 2. However, it was not clear if Stage 0 images were completely removed and we assumed this was the case given that Stage 0 versus Stage 1 is a challenging area due to the faint nature of the demarcation line.

Zhao et al., \cite{zhao2019deep} turned to deep learning for identifying Zone I using Retcam images.  Their approach is based on first identifying the position of the optic disc and macula. Thereafter, the center point of the optic disc and macula were calculated, from which Zone 1 was then calculated. This is the only study we know of that identifies Zone (I). They  achieve accuracy of 91\% with an IOU threshold of 0.8. They acknowledge that disc and macula positioning is required without which this study is incomplete. An automated Zone Quality Filter is a good contribution.

The above works showed promising results but had similar gaps. First challenge was with non-independently verified private ROP Retcam datasets. Next, dataset used were lacking geographical diversity to address this as a worldwide challenge. At time of writing, there was no public labelled ROP population diverse dataset which could be used to compare and measure results with, due to the sensitive nature of paediatric  data. A standardised open source or researchers only use,  ROP labelled dataset is critically needed by ROP AI researchers to create diagnostic solutions. It will allow for reproducibility of published work and aid future works. In terms of purist approach to Stages, and Zones With the exception of Ding et al., \cite{ding2020retinopathy} who focused on Stage only, others such as Tong et al., \cite{tong2020automated} combined various Stages and Plus disease for broader outcomes, similarly Wu et al., combined Stage and Zones minus Plus disease. Fundus pre-processing in all papers to date is very general except for Ding et al., \cite{ding2020retinopathy} and Mulay et al., \cite{mulay2019early} who used R-CNN to amplify the demarcation line. This motivated our study on researching improvements in the image pre-processing to assist in ROP classification. 

As part of our belief in open and collaborative research for the automation of ROP diagnosis using CNNs, Dr Kouroush and team created a labelled ROP repository which is available to use by fellow researchers with prior authorisation. This repository was used to support our focus on ROP Retcam pre-processing techniques that enhance overall features pertinent in providing high quality classifier outcomes in terms of Plus, Stages and Zones individually instead of grouping different ROP aspects. Data augmentation is used to supplement the lack of larger ROP dataset with transfer learning CNNs. We demonstrated that improving ROP features for ROP Retcam images using our improved pre-processing methods does result in improved deep learning classifier results. These methods can also aid other researchers who can re-try their experiments and provide their input on improvements using these methods.

\subsection{Fundus Pre-processing}
\label{sec:FundusPre-processing}
In the application of CNN to the ROP classification problem, we recognised that the crucial factor is the patients' Retcam images quality. Image pre-processing for paediatric Retcam captured images, a critical part of our work, is now reviewed. This can be defined as either the extraction, enhancing, or removal of features within the image. This includes pixel brightness and geometric transformations \cite{sonka2014image,zuiderveld1994contrast}. Methods are grouped either in image domain or restoration method categories. Most widely used methods fall in the first category and are popular primarily due to their simplicity. A second class of methods is categorised as restoration and are complex. The restoration class of methods are fairly new with novel techniques based on similar principles.  The former set of methods have primarily been used for ROP but the restoration category has not yet been documented in the improvement of ROP fundus image quality. To the best of our knowledge, this is the first time where restoration methods are used to improve the ROP Retcam quality. Described below are the primary methods used in both classes which are used and/or extended by this research.

\subsubsection{Image domain methods}
With ROP Retcam captured images, pre-processing is needed to normalize image brightness, correct for image non-uniformity, and reduce noise or image artefacts such that image clarity is restored. ROP Retcam images are challenged significantly by noise. This includes reflection, artifacts, and poor focus. Some conventional feature-based techniques for fundus pre-processing include Grayscale Conversion, Contrast Limiting Adaptive Histogram Equalization (CLAHE) \cite{zuiderveld1994contrast}, and Image filtering \cite{akram2009gabor,akram2014detection}. While there are other methods such as filtering, wavelet transformation, and morphological, we primarily focus on these mostly used methods for overall ROP image processing based on present literature survey.

A colour digital image contains Red, Green, Blue (RGB) channels with the green channel contains the most data.  First image processing method is Grayscale using green channel. This is performed to void any unique colour uniqueness that causes the model to pick additional characteristics resulting in its intensity information. It is implemented using the standard formula:

\begin{equation}
\textbf{I}o = ((0.3 \text{×} \textbf{R}) + (0.59 \text{×} \textbf{G}) + (0.11 \text{×} \textbf{B}))
\end{equation} 

The next three more related histogram based techniques are Histogram Equalization (HE) \cite{cheng2004simple,team_2023}, Adaptive Histogram Equalization (AHE) \cite{pizer1987adaptive} and Contrast Limited Adaptive Histogram Equalization (CLAHE) \cite{zuiderveld1994contrast}. Histogram Equalization (HE) is based on histogram calculated using image pixel intensity and then transforming it into a new histogram while Adaptive Histogram Equalization (AHE) calculates histograms by sections across the image and then distributes the brightness. Contrast Limited Adaptive Histogram Equalization (CLAHE) computes several histograms for different sections of the image, and subsequently distributes the lightness values but it caps the histogram to a predefined value to prevent over amplification that occurs with AHE. Resulting image is low noise sharpened which can assist in numerous medical diagnoses \cite{chaudhury2022effective,sheet2022retinal,jabbar2022transfer, gangwar2021diabetic}. CLAHE has achieved better results than the original Histogram Equalization (HE) \cite{cheng2004simple,team_2023}, and also better than Adaptive Histogram Equalization (AHE) \cite{pizer1987adaptive}. Consequently, CLAHE and Grayscale have been used in most ROP papers discussed earlier and for Diabetic Retinopathy classification \cite{jabbar2022transfer, gangwar2021diabetic} because of their simplicity and effectiveness. CLAHE's implementation also includes Lab color space, which provides another way to specify and quantify colour. 

Overall, the image domain set of methods are satisfactory but do not address the core challenge of restoration of ROP Retcam images adequately on their own. This is primarily due to haziness resulting from physical nature of multiple internal reflection as light passes through various layers and mediums of the eye's structure.

\subsubsection{Restoration methods}
When light enters the retina, it undergoes refraction prior to hitting the retina, and reflects from each layer it hits, including the sclera, lens, vitreous and retina itself.  This reflection contributes a significant amount of noise.  The formation of haze in the image is observed by the capturing fundus camera. The model for describing formation of haze is given in \cite{zhu2014single} as:

\begin{equation}
  \textbf{I}(x) = \textbf{J}(x) \textit{t}(x) + \textbf{A}(1- \textit{t}(x))
  \label{eqn:haze}
\end{equation}

where \textbf{I} is observed intensity, \textbf{J} is scene radiance, \textbf{A} is global atmospheric light, \textit{t} is the medium transmission describing the portion of light reaching the camera unscattered, and x is the pixel. The first term is known as direct attenuation and the second term is airlight \cite{he2010single}. We attempted, in our study, to leverage atmosphere based dehazing methods using Dark Prior Channel (DCP) as described first by He et al., \cite{he2010single} for each fundus image. DCP is based on the observation that in haze-free outdoor images, one of the channels will have a patch with the \textbf{A} variable having a low value in Equation~(\ref{eqn:haze}). Haze removal is performed by estimating the transmission, refining the transmission by soft matting, final scene radiance recovery, and finally estimating the atmospheric light as the highest intensity in the input image. While good for terrestrial use, this did not yield much success for ROP Retcam images. Similarly, other dehazing powerful techniques as proposed by Zhu et al., \cite{zhu2014single}, and Sami et al., \cite{sami2019novel} were also implemented without success in improving ROP Retcam images in terms of restoring details. The primary cause is these methods do not account for the fundamental challenge of reflection from within the eye.

For a hazy RGB image, neither channels will have non zero pixel values. Using amplification theory based on DCP theory, Gaudio et al., \cite{gaudio2020enhancement} proposed Pixel Color Amplification (PCA) which enhances a given fundus image. DCP theory permits inversions which can be used to derive additional Priors, namely Inverted DCP (Illumination Correction), and Bright Channel Prior (Exposure Correction). A fourth Prior based on the inter-relationship of three (3) priors is derived by Gaudio et al. \cite{gaudio2020enhancement}. Given Image \textbf{I}, transmission map \textbf{t} and atmosphere \textbf{A}, these four (4) Priors are unified with each revealing a weak and strong amplification including those of dark and bright pixel neighbourhoods. This yields 4 brightening/darkening methods which are referred to by letters A-D to brighten and W-Z to darken. These methods can be used individually or in combination to yield a merged image. Further a sharpening method is also available which can be activated by prefixing each letter with \textbf{s}. This allows sharpening of retinal features post prior computations which amplifies the difference between image and blurry computed version of itself. Using a combination of 4 Priors, PCA showed good retinal enhancement for EYEPACs/IRiD. It also held promise with our modification for ROP Retcam images that allows auto-balancing of illumination which is unpredictable. We noted that PCA suffered from being over-illuminated when used with adult fundus pictures as noted by Gaudio et al., \cite{gaudio2020enhancement}. The same problem was also noted when used with our ROP Retcam images dataset.

Zhang et al., \cite{zhang2022double} tackled the problem of reflection specific to a retina by building a multi-layer model of image formation that specifically dealt with reflective/illuminative imaging. The transmission term which He, Gaudio and others set to a constant value, is applied to illuminating light but also to reflected light. The first task is to estimate an enhanced restoration image value using illumination, transmission of lens and a scatter matrix. The second component restores the image by focusing on the retinal area ignoring the black exterior box. Coarse illumination correction is performed across red, green, and blue channels followed by fine illumination boosting where the Grayscale dark channel prior is used for dehazing. The last step is scatter suppression which results in an illumination corrected and dehazed image.  The method is called Double Pass Fundus Reflection (DPFR). This is very significant for ROP Retcam image pre-processing. It aligned with our own analysis of ROP Retcam images which notes that reflection from within the eye is the primary cause of quality degradation.

As the problem of ROP Retcam images is a multi-faceted challenge, we solved this by employing both image domain and restoration methods. First, we solved the problem of over/under amplification inherent in PCA while fine tuning DPFR specific for ROP Retcam images. These changes allow the pre-processed images to be more suitable for deep learning classifiers for ROP. We then combined our improved restoration methods and classic image domain method namely CLAHE to create a hybrid that improved the overall ROP features within the ROP Retcam image. Using each pre-processing method, a training and validation pre-processed dataset was created. Each pair was used with two different classifiers specific to Plus, Stages and Zones for ROP and results documented. 

\section{Methods}
\label{sec:DatasetsMethods}
\subsection{Background}
Image pre-processing is a critical step for enhancing Retcam as well as other captured fundus images which can either be in image domain or restoration methods. In this work, we first to solve Pixel Colour Amplification illumination problem in Section ~\ref{sec:PCAr}, and then describe the changes required in Double Pass Fundus Reflection for Retcam ROP images in Section ~\ref{sec:DPFRr}. In Section ~\ref{sec:EraseSegmentMap}, we highlight a segmentation based erosion method which can be used to remove blood vessels from a Retcam ROP image to further reduce the noise presented to a CNN by using segmentation. Lastly, 3 CNNs using transfer learnings were created to  classify pre-processed Retcam ROP images for Plus disease, Stages and Zones. The overall system we developed is refered to as McROP. 

\subsection{Datasets}
First datasets needed to be collected. Plus disease will either have Plus or No Plus. Stages will be labelled from Stage 0 to Stage 5.  Images for Zones will be labelled as Zones I, II or III. These are graded based on criteria as noted by in reference \cite{stephen2022} for the 3 seperate categories independently with the appropriate labels. 

A small ROP Retcam dataset. referred to as Calgary ROP dataset, was obtained as part of a collaboration agreement with University of Calgary with Dr Anna Ells's support. The Calgary ROP dataset had 1778 Retcam captured images with diagnosis ground truth. The original images were captured using Retcam camera in 640 $\times$ 480 pixels. Age, gender and race were not identified as per health data privacy requirements. Data quality guidelines included firstly identifying and excluding images which were badly focused, over illuminated, missing key retina features, having laser treatment marks or too many external reflections. Image of the exterior of the eyes were also excluded. This manual process highlighted a crucial gap in overall ROP data collection process, namely an automated CNN based Data Quality Filter tool. The remaining images were still challenged by poor illumination, and at times are slightly unfocused. The final dataset was re-checked again by Dr Kourosh Sabri as second grading specialist for final ground truth. 3 separate datasets were created. One each for Plus diseases (pre-Plus was excluded), Stages 0-3, and Zones I-III classification.

\subsection{Traditional image pre-processing methods } 
\label{NovelImgprocbackground}

Beginning with the image domain class models, there are three methods we introduced for our use.  These are Grayscale, CLAHE and our derived third method using a set of image domain class methods. Extensive use was made of the OpenCV library \cite{opencvgeo,opencvimgproc} for these methods. As discussed earlier, image domain pre-processing do not completely remediate the underlying problem but simply reduced it in different ways. 

\begin{enumerate}

    \item Grayscale (\textbf{Gray}) \cite{gonzalez2004digital} converts the colour to grayscale thereby voiding the image of any unique colour uniqueness. This prevents the CNN from identifying any colour based characteristics. 
     
    \item Contrast Limited Adaptive Histogram Equalization (\textbf{CLAHE}) \cite{zuiderveld1994contrast} is used to enhance the overall contrast of the image and smoothen out the pixels. A modified approach to CLAHE introduces a cliplimit of 2.0 across all three separated channels (Red, Green, and Blue) and merged into a single output. This gives a good contrast of ROP features.
         
    \item We derived another method by using CLAHE output and extracting the green channel only. Green channel always carries the highest level signal which aids in identifying the demarcation line visually while keeping other noise balanced. To this, Histogram Equalization is then applied for better contrasting. We refer to it as the CLAHE-Green-Histogram (\textbf{CGH}) method.

\end{enumerate}

The image domain methods do not address the underlying problem of ROP Retcam image quality, which emerges when the incidental light entering the retina is reflected back from not only the retina but the additional layers starting from the surface of the eye, lens, vitreous and retinal layer itself. After reviewing the haziness of the pictures, we deduced that this reflection problem is due to the close proximity of the capturing device laying on the surface of the retina and light induced into the eye. 

\subsection{Pixel Colour Amplification for ROP Retcam}
\label{sec:PCAr}
Pixel Colour Amplification (\textbf{PCA}) by Gaudio et al., \cite{gaudio2020enhancement} is an open-source image enhancement toolkit which uses 4 letters A-D to brighten and 4 letters W-Z to darken including sharpening. These can be combined to form an average. This can be controlled via choice of parameters. We extended this process to auto detect the illumination of original image and find corresponding 3 best results that is used to create a composite image. For ROP images, the resulting composite image, highlighted the demarcation line well as well as the retina blood vessels.  PCA pre-processed ROP Retcam images were challenged with over-saturation of reds in the center vision region as described by Dissopa et al., \cite{dissopa2021enhance}. This was significant challenge in ROP Retcam images when using PCA alone. In order to use it correctly, it was imperative to automatically balance the illumination and then contrast it further during post PCA pre-processed image. 

In this work, we solved the first PCA illumination problem by identifying its best output image in terms of illumination. First, the transmission maps \textbf{t} are obtained for each of the 4 methods. These are named as A,B,C,D for brightness, and W,X,Y,Z for darkness. Brightness and darkness is selected by altering the value of atmosphere \textbf{A}. Atmosphere value of 0 is for brightness and 1 for darkness. Final non-distorted image \textbf{J} is derived using each of the 8 distinct transmission maps \textbf{t}. This results in 8 output images produced for each input image. When deriving transmission map t, the depthmap is normalised such that using a median of 0.5 and returned a calculated score value relative to 0.5 for each method. In order not to over or under amplify, the image is centered and cropped to isolate region of interest map. This centre cropping is provided in IETK library. YCrCb \cite{sonka2014image} colour space was used for computing the depth map. This was chosen as its Y component captures the luminance/brightness with Cr and Cb that note the colour differences for red and blue.  Once all the solutions and their scores are obtained, record the scoring result with closest value to the median. The closest value to the median is the final selected balanced illuminated output as undistorted image \textbf{J}. The PCA Auto-Illumination algorithm we refer to as PCAr for ROP Retcam is described in Algorithm ~\ref{alg:Denoisepcar}.

\begin{algorithm}[H]
\caption {PCA Auto-Illumination (PCAr)}\label{alg:Denoisepcar}
\begin{algorithmic}
      \State \text{Input option for singleimage or compositeimage}
      \State \text{Load image (imageX) }
      \State \text{Crop and center the image for area of interest}
      \State \text{Normalise depthmap for solving \textbf{t} for median of 0.5}
	 \State \text{Evaluate all brightening/darkening functions A-D, W-Z}       
      \State \text{Set Atmosphere \textbf{A}=0 for brightness} 
      \State \text{Get solution of \textbf{J} for darkening methods for median pixel offset score:}
      \State \hspace{0.2cm} \text{A= \textbf{J}(imageX,\textbf{A},\textbf{t}(1-I)})
      \State \hspace{0.2cm} \text{B= \textbf{J}(imageX,\textbf{A},\textbf{t}(I)}) 
      \State \hspace{0.2cm} \text{C= \textbf{J}(imageX,\textbf{A},1-\textbf{t}(I)})
      \State \hspace{0.2cm} \text{D= \textbf{J}(imageX,\textbf{A},1-\textbf{t}(1-I)})
      \State \text{Set Atmosphere \textbf{A}=1 for darkness}  
      \State \text{Get solution of \textbf{J} for darkening methods for median pixel offset score:}
      \State \hspace{0.2cm} \text{W= \textbf{J}(imageX,\textbf{A},\textbf{t}(1-I)})
      \State \hspace{0.2cm} \text{X= \textbf{J}(imageX,\textbf{A},\textbf{t}(I)}) 
      \State \hspace{0.2cm} \text{Y= \textbf{J}(imageX,\textbf{A},1-\textbf{t}(I)}) 
      \State \hspace{0.2cm} \text{Z= \textbf{J}(imageX,\textbf{A},1-\textbf{t}(1-I)})

      \State \text{USING scored images A-D, W-Z}
      \State \hspace{0.20cm}\text{\textbf{IF} singleimage \textbf{THEN}}
      \State \hspace{0.35cm}\text{pcarimage = image with lowest score}
      \State \hspace{0.20cm}\text{\textbf{ELSIF} compositeimage \textbf{THEN}} 
      \State \hspace{0.35cm}\text{img1 = PCAr with lowest score in brightening group}
      \State \hspace{0.35cm}\text{img2 = PCAr with second score in brightening group}
      \State \hspace{0.35cm}\text{img3 = PCAr with highest score in darkening group}
      \State \hspace{0.35cm}\text{pcarimage = img1 + img2, img3}
      \State \hspace{0.20cm}\text{\textbf{ENDIF}}
      \State \text{Output pcarimage}

\end{algorithmic}
\end{algorithm}

\subsection{Double Pass Fundus Reflection for ROP Retcam}
\label{sec:DPFRr}
DPFR\cite{zhang2022double} is the latest novel technique which takes into account the image capturing mechanism which includes illumination, its forward journey into the eye and reflection. The transmission index is also estimated and not fixed as in Pixel Colour Amplification (PCA). The final image S is computed using the sum of two back scattering components \textbf{R}{\tiny o}  from retina and \textbf{R}{\tiny 1} being from intra-ocular.


\begin{equation}
\textbf{S}(r) = \textbf{R}{\scriptstyle\ o} + \textbf{R}{\scriptstyle\ 1}=\textbf{I}{\scriptstyle\ ill} (r).\textbf{T}^2{\scriptstyle\ lens}(r).[{\textbf{T}^2{\scriptstyle\ sc}}(r).\textbf{O}(r)+1 \text{-} \textbf{T}{\scriptstyle\ sc}(r)]
\label{eqn:dpfr}
\end{equation}

Restoration of image O is implied if the illumination matrix \textbf{I}{\tiny ill} and transmission matrix \textbf{T} lens and \textbf{T}{\tiny sc} can be measured using ‘prior’ which is pre-known information. By estimating \textbf{I}{\tiny ill} , \textbf{T}{\tiny lens} and \textbf{T}{\tiny sc} and solving Equation~\ref{eqn:dpfr}, an enhanced version is \textbf{O} obtained.  The original method has image pre-processing, coarse followed by fine illumination, and scatter suppression.  The resulting image is very grainy when viewed closely for ROP classification use especially for Stages. In this work, We enhanced DPFR further for make it suitable for ROP Retcam images by altering three method values. Firstly, Low pass filter value of $\epsilon$ was changed in Coarse Illumination Correction method to a lower value. Next, dehazing estimate target method, the estimate for dehazing was doubled. Lastly, in Fine Illumination method, the estimate for dehazing was set to a much lower value. This revision we note this as a slightly altered method denoted as \textbf{DPFRr} tuned for ROP Retcam images. The changes also reduce the level of choroid vessels. 

Two further avenues were explored to reduction of choroid vessels. First with the application of CLAHE and denoising ROP Retcam image using wavelets. By applying CLAHE to three channels prior to remerging it,  DPFRr output image's contrast can be reduced considerably yielding the presence and features of demarcation line. Its positive side effect is the removal of any additional colouration which can influence the classifier. We refer to this hybrid as \textbf{DPFRr-CLAHE} as additional improved method. The second alternative to CLAHE using output of DPFRr was also examined. We attempted to denoise the Retcam image using wavelets combined with guided filter. Application of wavelets resulted in significant reduction of blood vessels including fovea in green channel but did not eliminate them completely. A side effect included the reduced visibility of the faint Stage 1 demarcation line. We note the later as a possible avenue for future work.

\subsection{Erosion of blood vessels using Segmentation map}
\label{sec:EraseSegmentMap}

Having improved the quality of the ROP Retcam images for Stage classification, we further explored the elimination of the blood vessels using segmentation. The aim was to reduce the noise a classifier has to contend with. Noise from choroid vessels for a premature infant is very significant in an ROP Retcam. For Stage diagnosis, when using the full ROP Retcam image, non essential retinal features such as optic disc, and blood vessels are unnecessary for the classifier to consider. The objective was reduction of noise from blood vessels by eroding it out of view. For this, we turned to U-Net \cite{ronneberger2015u} segmentation as already there is a large body of knowledge without diverging from our core work to leverage from. The choice was made of LWNET \cite{galdran2022state} which is a minimal model segmentation design. LWNET has ability to generate both coarse and fine vessel segmentation. LWNET has been pre-trained using open source retina fundus images to generate vessel maps which have been pre-validated against hand generated fundus maps of adult fundus images.

\begin{figure}[h]
    \centering
    \includegraphics[height=3in,width=4in]{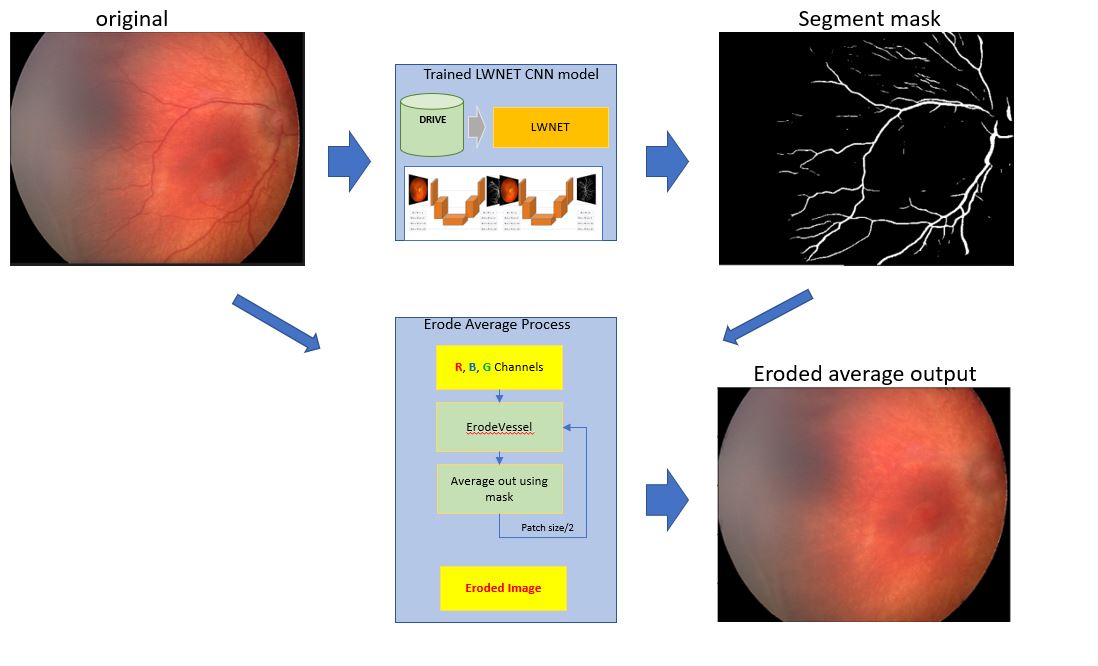}
    \caption{Erosion of blood vessels using LWNET vessel map}
    \label{fig:FundusLWNETErase}
\end{figure}

We used the coarse segmentation for larger vessels as proof of concept. Using pre-trained segmentation model of LWNET, we passed both the original training and validation Stage datasets to create respective segment maps. Next, an new algorithm was created to leverage this segmentation map inversely to erase corresponding blood vasculature pixels using choice of average colour of adjacent region starting with coarse square of 32 and then scaling downwards to 4 pixels as noted in Algorithm 2. Alternatively method available is to substitute the average using a gaussian kernel. The process is described in Figure \ref{fig:FundusLWNETErase}. On the original colour ROP Retcam image, this was able to erode using average kernel or reduce the intensity using gaussian kernel. This greatly helped focus the image towards the demarcation line.

\begin{algorithm}[H]
\caption {Erosion using Segment masking}\label{Erosion}
 \begin{algorithmic}

   \Function{cleanimage}{$imageX, maskY$}
      \State split into 3 channels and employ gaussian to the vessel using mask
      \State r = blendvessel(imageX(r channel), maskY)
      \State g = blendvessel(imageX(g channel), maskY)
      \State b = blendvessel(imageX(b channel), maskY)
      \State \Return \textit{stackedimage(r,g,b)}
    \EndFunction

	\Function {chooseside}{$boxareaV, channelV, maskV$}
       \State boxarea = boxareaV.flatten()
       \State channel = channelV.flatten()
       \State mask = maskV.flatten()
    
       \State \textit{mask will have veins segmented map with values greater than 0.1}
       \State veins = (mask greater than 0.1)
       
       \State retval = np.zeros(boxarea.shape)
       \State \textit{assign gaussian to selected pixels}
       \State retval[veins] $=$ boxarea[veins]
       \State \textit{ keep remaining ones same as original}\
       \State {$retval[\sim\ veins] = channel[\sim\ veins]$}
       
       \State \Return \textit {retval.reshape(channelv.shape)}
    \EndFunction

   \Function {blendvessel}{$channel,mask, psize, gaussianopt$}
       \State \textit{-- removes vessel from each channel using gaussian or average}
       \State \textit{-- psize default of 32 or any other value passed}
       
	   \State patchsize = psize
          \State \textbf{loop} $while patchsize ~{\geq}~ 2$ \ 

     \State \hspace{0.2cm}{blursqr = convolve2d(channel, np.one((patchsize,patchsize)) using wrap)}
		\State\hspace{0.2cm}\textbf{if} guassianopt \textbf{then} blursqr = gaussian(blursqr)
  	\State\hspace{0.2cm}\textbf{else} blursqr = average(blursqr)

	    \State\hspace{0.2cm}curedchannel = chooseside(blursqr, channel, mask) 
     \State \hspace{0.2cm}channel = curedchannel	
        \State \hspace{0.2cm}patchsize = patchsize/2
        \State \textbf{end loop}
        \State \Return \textit {curedchannel}
    \EndFunction
\end{algorithmic}
\end{algorithm}


\subsection{ROP CNNs using Transfer Learning }

Transfer learning \cite{developerguide} has been demonstrated with repeated success to resolve challenges in deep learning CNNs by trying to solve a problem similar to what it had been trained for. It has been used successfully to improve classifiers \cite{sheet2022retinal,jabbar2022transfer,gangwar2021diabetic}. As this project had limited number of ROP images, transfer learning is a critical component as it cuts down the time required for training and tuning by using accumulated learnings from prior datasets towards a new problem. This is advantageous as recognizing patterns, shapes etc are basis for all models irrespective of classification. In this project, We used ResNet50, InceptionResv2 and Dense169 \cite{resnet50,inceptionresv2,densenet} which were already pre-trained on ImageNet dataset and provided by Tensorflow/Keras \cite{developerguide}. These networks ideally suited as they are pre-optimised for identity mapping during the training process. In all cases, we do not want the last fully connected layer. This is replaced by our task for Plus, Stage and Zone. We freeze the weights of each model to ``False '' to stop updates to the pretrained weights, as we want to preserve knowledge retained earlier using the ImageNet dataset. We add a custom block of CNN layers on top of each one of these models to build a hybrid model.  It is helpful when there is not sufficient data to train a full model, reduces overfitting and assists in the training process. The shape of each pretrained model’s input layer and final dense layer is modified. Weights are additionally calculated based on the number of training images to prevent overfitting. Additionally, L1 regularizer and Dropout were also employed to reduce overfitting. Use was made of GradCam\cite{gradcam} to check for features being considered. This is shown in Fig 2 under Convolution Neural Networks block diagram.

We describe layers for each of the CNNs created.  For Plus, Stages, and Zones, in ResNet50, following layers were frozen and additional ones added with softmax given below:

\begin{algorithm}[H]
\caption {ResNet50 Transfer Learning model - Plus, Stage, Zones}\label{ResNet50}
 \begin{algorithmic}[1]
      \State Freeze all layers except: 'res5c\_branch2b', 'res5c\_branch2c', 'activation\_97'
      \State Add fully connected layers:\
      	   \newline GlobalMaxPooling2D\
      	   \newline Dense(1024,'relu',l1)\
      	   \newline Dropout(0.2) \
      	   \newline Flatten\
      	   \newline Dense(1024,'relu','l1')\
      	   \newline Dropout(0.2)\
      	   \newline Batchnormalization\     	
      \State Final classifier prediction using softmax activation function
\end{algorithmic}
\end{algorithm}

Following similar practise, we extend InceptionResv2 for Plus in following summary which is slightly different than for Stage and Zones in subsequent definition.

\begin{algorithm}[H]
\caption {InceptionResv2 Transfer Learning model - Plus}\label{InceptionResv2-1}
 \begin{algorithmic}[1]
      \State Freeze all layers except: 'block8\_10\_mixed' that is trainable
      \State Add fully connected layers:\
		   \newline Flatten \
		   \newline Dense(1024, 'relu’)\
		   \newline Dropout(x) where x=0.6 for Plus\
		   \newline Dense(1024,'relu')\
		   \newline Dropout(x) where x=0.6 for Plus\
		   \newline Batchnormalization\     	
      \State Final classifier prediction using softmax activation function
\end{algorithmic}
\end{algorithm}

\begin{algorithm}[H]
\caption {InceptionResv2 Transfer Learning model - Stage, Zones}\label{InceptionResv2-2}
 \begin{algorithmic}[1]
      \State Freeze all layers except: 'block8\_10\_mixed' that is trainable
      \State Add fully connected layers:\
		   \newline Flatten \
		   \newline BatchNormalisation\
		   \newline Dropout(x) where x=0.7 for Stage and 0.2 for Zones\
		   \newline Dense(1024,'relu', l1)\
		   \newline Dropout\
		   \newline Dense(1024,relu','l1')\
		   \newline Dropout(x) where x=0.7 for Stage and 0.2 for Zones\
		   \newline Batchnormalization\     	
      \State Final classifier prediction using softmax activation function
\end{algorithmic}
\end{algorithm}

Lastly, Dense169 was used for Stages to cross validate results between ResNet50 and InceptionResv2 for Stages.

\begin{algorithm}[H]
\caption {Dense169 Transfer Learning model - Stages}\label{Dense169-1}
 \begin{algorithmic}[1]
      \State Freeze all layers except: 'block8\_10\_mixed' that is trainable
      \State Add fully connected layers:\
		   \newline GlobalMaxPooling2D() \
		   \newline Dense(1024, 'relu’, 'l1')\
		   \newline Dropout(x) where x=0.2\
		   \newline Dense(1024,'relu', 'l1')\
		   \newline Dropout(x) where x=0.2\
		   \newline Batchnormalization\     	
      \State Final classifier prediction using softmax activation function
\end{algorithmic}
\end{algorithm}

In the above subsections, we have presented the 3 core image pre-processing contributions and leveraging transfer learning based classifiers using ResNet50, InceptionResv2, and DenseNet for training/validation. These use prepared datasets that have been either pre-processed using different methods or baseline unprocessed Retcam images.  We next discuss the experiments. 

\section{Results}
\label{sec:Results}

The results from our system for each classifier and method of pre-processing are discussed next. Noting the limited dataset, by extensively using transfer learning with augmented training datasets, the classifiers demonstrated improvements resulting from the use of the two improved novel image pre-processors in comparison to traditional methods. These results for each pre-processing method by classifiers are shown separately in terms of Plus disease, Stages, and Zones. We discuss the results and compare them to respective peer research papers followed by further analysis.

\subsection{Pre-processing Methods outputs}

With data collection completed, labelling and split into training and validation datasets, image pre-processing stage was next performed.  Using this original pair, seven (7) additional image pre-processing algorithms were used to generate pairs of pre-processed training/validation datasets. All the resulting images still remained in 640x480 format. A total of 8 paired standardized datasets (1 original, 7 pre-processed) were ready for training and validation. 

\begin{figure}[H]
\caption{McROP Retcam samples using image domain based Pre-processing}
\centering
\includegraphics[height=4in,width=3.5in]{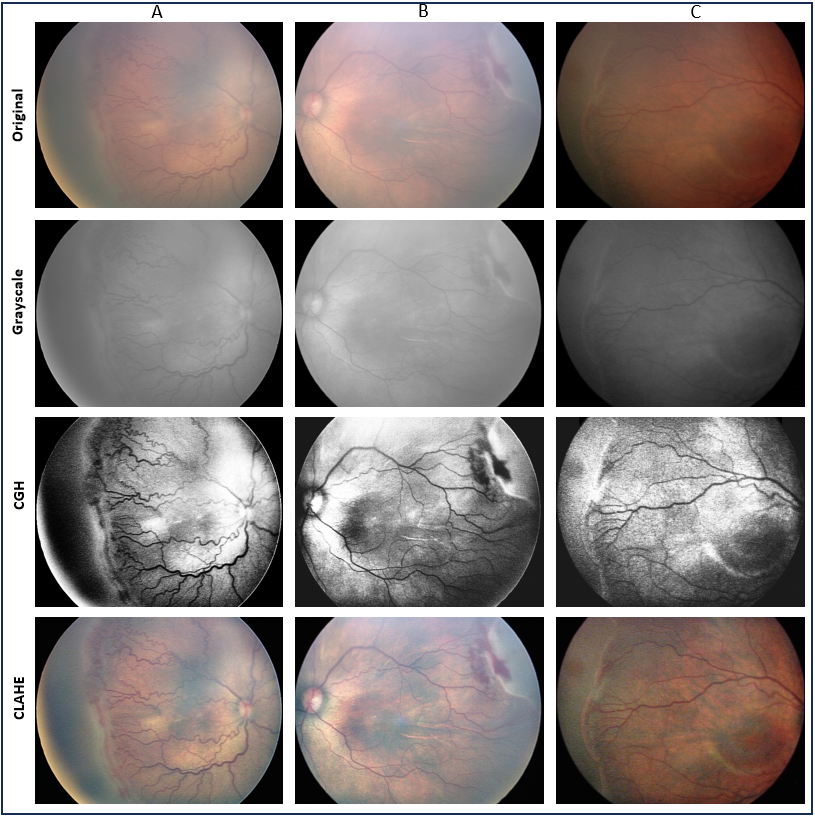}
 \label{fig:imagedomainpreproc}
\end{figure}

We first reviewed the image pre-processing outputs to summarise their key features. In the case of image domain methods, grayscale images demonstrated even shade of gray with brightness evenly distributed with a low level contrast. This method has a degree of reflection present that caused inability to highlight ROP specific features. CGH method further improved on grayscale and provided a significantly better contrast for blood vessels. The demarcation line was also very visible including macula. 3 channel CLAHE colour output showed better colour contrast and sharper output than the original unprocessed image. This allowed good visualisation of ROP features. Figure ~\ref{fig:imagedomainpreproc} shows 3 ROP Retcam original sample images from McROP dataset are shown with respective grayscale, CGH and CLAHE results underneath the original. 

\begin{figure}[H]
\caption{McROP Retcam samples using restoration based based Pre-processing}
\centering
\includegraphics[height=4.3in,width=3.5in]{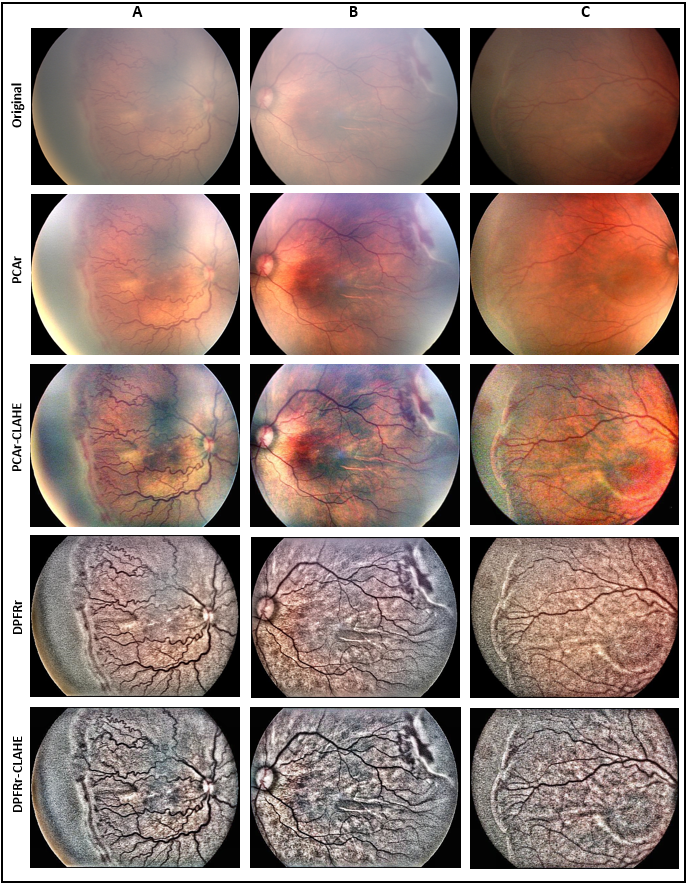}
 \label{fig:restorationpreproc}
\end{figure}

In the case of the improved restoration methods, PCAr composites showed better contrasting than CLAHE. ROP features were visible, but there were challenges with red colour being amplified from the original image. This was resolved with the application of 3 channel CLAHE. The result of PCAr-CLAHE firstly reduced overall reflection and improved the contrast.  Blood vessels, macula, optic disc as well as demarcation line became very pronounced in terms of visibility. DPFRr for same original shows a remarkable cancellation of the reflection making the vessels more pronounced.  This was much improved in comparison to PCAr-CLAHE.  However reddish colour remains speckled across the image. By applying 3 channel CLAHE to DPFRr in DPRFr-CLAHE method, it cancelled this problem and presented an image where ROP features were sharper and pronounced. DPFRr-CLAHE as well as PCAr-CLAHE both allowed for good visualisation of ROP features. Figure ~\ref{fig:restorationpreproc} shows 3 ROP Retcam original sample images from McROP dataset are shown with respective PCAr, PCAr-CLAHE, DPRFr, and DPFRr-CLAHE results underneath the original. 

\begin{figure}[H]
\caption{McROP Retcam samples using Erosion based Pre-processing}
\centering
\includegraphics[height=3.5in,width=3.5in]{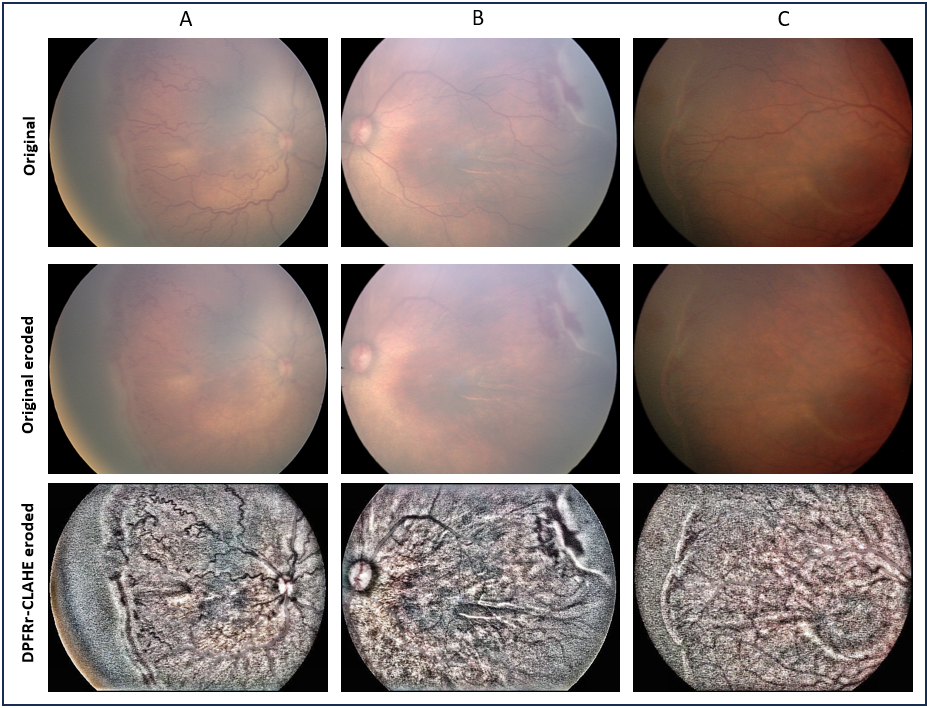}
 \label{fig:erodepreproc}
\end{figure}

We had also prepared eroded datasets .  The first dataset was the original stages McROP dataset.  We applied the erosion algorithm ~\ref{Erosion}. The eroded original images show a complete erosion based off the generated maps. This image was then applied with DPFRr-CLAHE.  The result showed the presence of the background where the vessels had been eroded successfully in regular image.  However, in DPRFr-CLAHE where reflection and colour contrasting has been resolved, there is visiblity of eroded lines which can challenge the classifier with DPFRr-CLAHE. Figure ~\ref{fig:erodepreproc} shows 3 ROP Retcam original sample images from McROP dataset with respective original eroded images and DPFRr-CLAHE output of the eroded original image underneath respective original images. From these results, we recommend using alternative approach such as Mask R-CNN segmentation for Stages detection. Mask R-CNN will leverage a boxed demarcation zone and extracts it out. In addition, the training image for Mask R-CNN can be further denoised using wavelets to improve the detection detect and box the demarcation line. The erosion approach using segmentation can be utilised in clinical settings if other pathologies need to be viewed without vessel noise.

We were able to see improvements in overall ROP Retcam image quality by employing image pre-processing.  The best outputs were from PCAr-CLAHE and DPFRr-CLAHE which provided significantly improved ROP features in comparison to other methods. With the pre-processing methods in place, we discuss the results of each of these methods in terms of ROP Classification using CNNs in the next sections. 

\subsection{Plus Disease Classification}

Using Plus/No Plus datasets, we trained and validated with Resnet50 and InceptionResv2 CNN classifiers for each image pre-processing methods as well as with unprocessed images. In all cases, same original dataset pair was the baseline with which 7 pre-processed dataset pairs were generated. This ensured direct comparison. The results obtained for each classifier trained independently for paired dataset is illustrated in Tables ~\ref{table:TabledatasetsPlusresnet}, and~\ref{table:TabledatasetsB1Plus}. The best results are highlighted in yellow/bold.  In ResNet50 CNN, for Plus disease, using the same data but pre-processed showed slight improvements based on Sensitivity, Specificity, Precision, and F1 Score. Both DPFR/DPFRr-CLAHE and showed best results over other methods with PCAr being tied as second best. With InceptionResv2, similar results to ResNet50 were found for Plus disease. Unprocessed (baseline) had lowest sensitivity with DPFRr and DPFRr-CLAHE showed the best results similar to ResNet50. PCAr/PCAr-CLAHE did not perform well here. With InceptionResv2, we are able to see a clear trend of improvements with each type of pre-processing method starting from grayscale and reach best outcome using restoration method of DPFRr/DPFRr-CLAHE. This improvement can be directly attributed by the improved visibility of blood vessels required for this classification.

\begin{table}[!ht]
\centering
\scalebox{0.5}{
\resizebox{\textwidth}{!}{
\begin{tabular}{l|llllll|}
\cline{2-7}
 &
  \multicolumn{6}{c|}{\cellcolor[HTML]{DDEBF7}\textbf{ResNet50 - Plus Disease}} \\ \hline
\rowcolor[HTML]{E7E6E6} 
\multicolumn{1}{|l|}{\cellcolor[HTML]{E7E6E6}\textbf{Methods}} &
  \multicolumn{1}{c|}{\cellcolor[HTML]{E7E6E6}\textbf{Sensitivity}} &
  \multicolumn{1}{c|}{\cellcolor[HTML]{E7E6E6}\textbf{Specificity}} &
  \multicolumn{1}{c|}{\cellcolor[HTML]{E7E6E6}\textbf{Precision}} &
  \multicolumn{1}{c|}{\cellcolor[HTML]{E7E6E6}\textbf{F1}} &
  \multicolumn{1}{c|}{\cellcolor[HTML]{E7E6E6}\textbf{Kappa}} &
  \multicolumn{1}{c|}{\cellcolor[HTML]{E7E6E6}\textbf{Accuracy}} \\ \hline
\rowcolor[HTML]{DDEBF7} 
\multicolumn{1}{|l|}{\cellcolor[HTML]{DDEBF7}Base} &
  \multicolumn{1}{l|}{\cellcolor[HTML]{DDEBF7}0.9764} &
  \multicolumn{1}{l|}{\cellcolor[HTML]{DDEBF7}0.9764} &
  \multicolumn{1}{l|}{\cellcolor[HTML]{DDEBF7}0.9764} &
  \multicolumn{1}{l|}{\cellcolor[HTML]{DDEBF7}0.9764} &
  \multicolumn{1}{l|}{\cellcolor[HTML]{DDEBF7}0.9529} &
  0.9764 \\ \hline
\multicolumn{1}{|l|}{Gray} &
  \multicolumn{1}{l|}{0.9764} &
  \multicolumn{1}{l|}{0.9764} &
  \multicolumn{1}{l|}{0.9764} &
  \multicolumn{1}{l|}{0.9764} &
  \multicolumn{1}{l|}{0.9529} &
  0.9764 \\ \hline
\rowcolor[HTML]{DDEBF7} 
\multicolumn{1}{|l|}{\cellcolor[HTML]{DDEBF7}CLAHE} &
  \multicolumn{1}{l|}{\cellcolor[HTML]{DDEBF7}0.9764} &
  \multicolumn{1}{l|}{\cellcolor[HTML]{DDEBF7}0.9764} &
  \multicolumn{1}{l|}{\cellcolor[HTML]{DDEBF7}0.9764} &
  \multicolumn{1}{l|}{\cellcolor[HTML]{DDEBF7}0.9764} &
  \multicolumn{1}{l|}{\cellcolor[HTML]{DDEBF7}0.9529} &
  0.9764 \\ \hline
\multicolumn{1}{|l|}{CGH} &
  \multicolumn{1}{l|}{0.9647} &
  \multicolumn{1}{l|}{0.9647} &
  \multicolumn{1}{l|}{0.9647} &
  \multicolumn{1}{l|}{0.9647} &
  \multicolumn{1}{l|}{0.9294} &
  0.9647 \\ \hline
\rowcolor[HTML]{DDEBF7} 
\multicolumn{1}{|l|}{\cellcolor[HTML]{DDEBF7}DPFRr} &
  \multicolumn{1}{l|}{\cellcolor[HTML]{DDEBF7}0.9764} &
  \multicolumn{1}{l|}{\cellcolor[HTML]{DDEBF7}0.9764} &
  \multicolumn{1}{l|}{\cellcolor[HTML]{DDEBF7}0.9764} &
  \multicolumn{1}{l|}{\cellcolor[HTML]{DDEBF7}0.9529} &
  \multicolumn{1}{l|}{\cellcolor[HTML]{DDEBF7}0.9529} &
  0.9764 \\ \hline
\rowcolor[HTML]{FFFF00} 
\multicolumn{1}{|l|}{\cellcolor[HTML]{FFFF00}\textbf{DPFRr\_CLAHE}} &
  \multicolumn{1}{l|}{\cellcolor[HTML]{FFFF00}\textbf{0.9764}} &
  \multicolumn{1}{l|}{\cellcolor[HTML]{FFFF00}\textbf{1.0000}} &
  \multicolumn{1}{l|}{\cellcolor[HTML]{FFFF00}\textbf{1.0000}} &
  \multicolumn{1}{l|}{\cellcolor[HTML]{FFFF00}\textbf{0.9764}} &
  \multicolumn{1}{l|}{\cellcolor[HTML]{FFFF00}0.9529} &
  \textbf{0.9764} \\ \hline
\rowcolor[HTML]{DDEBF7} 
\multicolumn{1}{|l|}{\cellcolor[HTML]{DDEBF7}PCAr} &
  \multicolumn{1}{l|}{\cellcolor[HTML]{DDEBF7}0.9764} &
  \multicolumn{1}{l|}{\cellcolor[HTML]{DDEBF7}0.9764} &
  \multicolumn{1}{l|}{\cellcolor[HTML]{DDEBF7}0.9764} &
  \multicolumn{1}{l|}{\cellcolor[HTML]{DDEBF7}0.9529} &
  \multicolumn{1}{l|}{\cellcolor[HTML]{DDEBF7}0.8588} &
  0.9764 \\ \hline
\multicolumn{1}{|l|}{PCAr\_CLAHE} &
  \multicolumn{1}{l|}{0.9351} &
  \multicolumn{1}{l|}{1.0000} &
  \multicolumn{1}{l|}{1.0000} &
  \multicolumn{1}{l|}{0.9664} &
  \multicolumn{1}{l|}{0.8824} &
  0.9412 \\ \hline
\end{tabular}%
}
}
\caption{ResNet50 results for Plus disease for each pre-processing method }
\label{table:TabledatasetsPlusresnet}
\end{table}

\begin{table}[!ht]
\centering
\scalebox{0.5}{
\resizebox{\textwidth}{!}{%
\begin{tabular}{l|llllll|}
\cline{2-7}
 &
  \multicolumn{6}{c|}{\cellcolor[HTML]{DDEBF7}\textbf{InceptionResv2 - Plus Disease}} \\ \hline
\rowcolor[HTML]{E7E6E6} 
\multicolumn{1}{|c|}{\cellcolor[HTML]{E7E6E6}\textbf{Methods}} &
  \multicolumn{1}{c|}{\cellcolor[HTML]{E7E6E6}\textbf{Sensitivity}} &
  \multicolumn{1}{c|}{\cellcolor[HTML]{E7E6E6}\textbf{Specificity}} &
  \multicolumn{1}{c|}{\cellcolor[HTML]{E7E6E6}\textbf{Precision}} &
  \multicolumn{1}{c|}{\cellcolor[HTML]{E7E6E6}\textbf{F1}} &
  \multicolumn{1}{c|}{\cellcolor[HTML]{E7E6E6}\textbf{Kappa}} &
  \multicolumn{1}{c|}{\cellcolor[HTML]{E7E6E6}\textbf{Accuracy}} \\ \hline
\rowcolor[HTML]{DDEBF7} 
\multicolumn{1}{|l|}{\cellcolor[HTML]{DDEBF7}Base} &
  \multicolumn{1}{l|}{\cellcolor[HTML]{DDEBF7}0.9111} &
  \multicolumn{1}{l|}{\cellcolor[HTML]{DDEBF7}1.0000} &
  \multicolumn{1}{l|}{\cellcolor[HTML]{DDEBF7}1.0000} &
  \multicolumn{1}{l|}{\cellcolor[HTML]{DDEBF7}0.9536} &
  \multicolumn{1}{l|}{\cellcolor[HTML]{DDEBF7}0.8353} &
  0.9176 \\ \hline
\multicolumn{1}{|l|}{Gray} &
  \multicolumn{1}{l|}{0.9474} &
  \multicolumn{1}{l|}{1.0000} &
  \multicolumn{1}{l|}{1.0000} &
  \multicolumn{1}{l|}{0.9729} &
  \multicolumn{1}{l|}{0.9059} &
  0.9529 \\ \hline
\rowcolor[HTML]{DDEBF7} 
\multicolumn{1}{|l|}{\cellcolor[HTML]{DDEBF7}CLAHE} &
  \multicolumn{1}{l|}{\cellcolor[HTML]{DDEBF7}0.9474} &
  \multicolumn{1}{l|}{\cellcolor[HTML]{DDEBF7}1.0000} &
  \multicolumn{1}{l|}{\cellcolor[HTML]{DDEBF7}1.0000} &
  \multicolumn{1}{l|}{\cellcolor[HTML]{DDEBF7}0.9729} &
  \multicolumn{1}{l|}{\cellcolor[HTML]{DDEBF7}0.9059} &
  0.9529 \\ \hline
\multicolumn{1}{|l|}{CGH} &
  \multicolumn{1}{l|}{0.9333} &
  \multicolumn{1}{l|}{0.8000} &
  \multicolumn{1}{l|}{0.9722} &
  \multicolumn{1}{l|}{0.9524} &
  \multicolumn{1}{l|}{0.8353} &
  0.9176 \\ \hline
\rowcolor[HTML]{FFFF00} 
\multicolumn{1}{|l|}{\cellcolor[HTML]{FFFF00}\textbf{DPFRr}} &
  \multicolumn{1}{l|}{\cellcolor[HTML]{FFFF00}\textbf{0.9722}} &
  \multicolumn{1}{l|}{\cellcolor[HTML]{FFFF00}\textbf{1.0000}} &
  \multicolumn{1}{l|}{\cellcolor[HTML]{FFFF00}\textbf{1.0000}} &
  \multicolumn{1}{l|}{\cellcolor[HTML]{FFFF00}\textbf{0.9859}} &
  \multicolumn{1}{l|}{\cellcolor[HTML]{FFFF00}\textbf{0.9529}} &
  \textbf{0.9759} \\ \hline
\rowcolor[HTML]{FFFF00} 
\multicolumn{1}{|l|}{\cellcolor[HTML]{FFFF00}\textbf{DPFRr\_CLAHE}} &
  \multicolumn{1}{l|}{\cellcolor[HTML]{FFFF00}\textbf{0.9722}} &
  \multicolumn{1}{l|}{\cellcolor[HTML]{FFFF00}\textbf{1.0000}} &
  \multicolumn{1}{l|}{\cellcolor[HTML]{FFFF00}\textbf{1.0000}} &
  \multicolumn{1}{l|}{\cellcolor[HTML]{FFFF00}\textbf{0.9859}} &
  \multicolumn{1}{l|}{\cellcolor[HTML]{FFFF00}\textbf{0.9529}} &
  \textbf{0.9759} \\ \hline
\rowcolor[HTML]{DDEBF7} 
\multicolumn{1}{|l|}{\cellcolor[HTML]{DDEBF7}PCAr} &
  \multicolumn{1}{l|}{\cellcolor[HTML]{DDEBF7}0.9113} &
  \multicolumn{1}{l|}{\cellcolor[HTML]{DDEBF7}1.0000} &
  \multicolumn{1}{l|}{\cellcolor[HTML]{DDEBF7}1.0000} &
  \multicolumn{1}{l|}{\cellcolor[HTML]{DDEBF7}0.9536} &
  \multicolumn{1}{l|}{\cellcolor[HTML]{DDEBF7}0.8352} &
  0.9176 \\ \hline
\multicolumn{1}{|l|}{PCAr\_CLAHE} &
  \multicolumn{1}{l|}{0.9113} &
  \multicolumn{1}{l|}{1.0000} &
  \multicolumn{1}{l|}{1.0000} &
  \multicolumn{1}{l|}{0.9536} &
  \multicolumn{1}{l|}{0.8352} &
  0.9176 \\ \hline
\end{tabular}%
}
}
\caption{InceptionResv2 results for Plus disease for each pre-processing method }
\label{table:TabledatasetsB1Plus}
\end{table}

\begin{table}[!ht]
\resizebox{\columnwidth}{!}{%
\begin{tabular}{l|llllll|}
\cline{2-7}
\textbf{} &
  \multicolumn{6}{l|}{\cellcolor[HTML]{DDEBF7}\textbf{Performance   of DPRFr-InceptionResv2 Architecture - Plus Disease}} \\ \hline
\rowcolor[HTML]{E7E6E6} 
\multicolumn{1}{|l|}{\cellcolor[HTML]{E7E6E6}\textbf{Type}} &
  \multicolumn{1}{l|}{\cellcolor[HTML]{E7E6E6}\textbf{Features}} &
  \multicolumn{1}{l|}{\cellcolor[HTML]{E7E6E6}\textbf{Sensitivity}} &
  \multicolumn{1}{l|}{\cellcolor[HTML]{E7E6E6}\textbf{Specificity}} &
  \multicolumn{1}{l|}{\cellcolor[HTML]{E7E6E6}\textbf{Precision}} &
  \multicolumn{1}{l|}{\cellcolor[HTML]{E7E6E6}\textbf{F1 Score}} &
  \textbf{Accuracy} \\ \hline
\multicolumn{1}{|l|}{\cellcolor[HTML]{DDEBF7}\textbf{No Plus}} &
  \multicolumn{1}{l|}{Normal} &
  \multicolumn{1}{l|}{0.9700} &
  \multicolumn{1}{l|}{1.0000} &
  \multicolumn{1}{l|}{1.0000} &
  \multicolumn{1}{l|}{0.9900} &
  0.9765 \\ \hline
\multicolumn{1}{|l|}{\cellcolor[HTML]{DDEBF7}\textbf{Plus}} &
  \multicolumn{1}{l|}{Presence of Plus disease} &
  \multicolumn{1}{l|}{1.0000} &
  \multicolumn{1}{l|}{0.97} &
  \multicolumn{1}{l|}{0.85} &
  \multicolumn{1}{l|}{0.9200} &
  0.9765 \\ \hline
\end{tabular}%
}
\caption{Plus Classification breakdown for DPFRr-CLAHE-ResNet50 }
\label{table:TabledatasetsPlusResNet50}
\end{table}

\begin{table}[!ht]
\begin{tabular}{|l|ll|}
\hline
                                         & \multicolumn{2}{l|}{\cellcolor[HTML]{DAE8FC}\textbf{Confusion Matrix}}                     \\ \hline
\rowcolor[HTML]{DAE8FC} 
\textbf{Plus Disease}                    & \multicolumn{1}{l|}{\cellcolor[HTML]{DAE8FC}\textbf{No Plus}} & \textbf{Plus}              \\ \hline
\cellcolor[HTML]{DAE8FC}\textbf{No Plus} & \multicolumn{1}{l|}{\cellcolor[HTML]{ECF4FF}72}               & 0                          \\ \hline
\cellcolor[HTML]{DAE8FC}\textbf{Plus}    & \multicolumn{1}{l|}{2}                                        & \cellcolor[HTML]{ECF4FF}11 \\ \hline
\end{tabular}
\caption{Confusion Matrix for Plus Disease for DPFRr-CLAHE}
\label{table:TabledatasetsC1Plus}
\end{table}

The best overall image pre-processing result for Plus disease detection was evenly split between ResNet50 and InceptionResv2 CNNs with DPFRr-CLAHE pre-processing method providing best sensitivity/specificity. Using the ResNet50/DPFRr-CLAHE as the best of the three top, the result was then further broken down by No Plus/Plus disease in Fig ~\ref{table:TabledatasetsPlusResNet50} using confusion matrix in Table~\ref{table:TabledatasetsC1Plus} that resulted in overall accuracy of 97.65\% which is marginally better than InceptionResv2/DPFR-CLAHE's accuracy of 98.59\%. As both ResNet50 and InceptionResv2 for this method gave best results, we chose ResNet50 architecture. ResNet50 with DPFRr-CLAHE yielded the best results overall with sensitivity and specificity of 97.64\% and 100\%. Precision was 100\%.  This was as of the point of writing, the only research where Plus disease classification had been performed using a restoration based image pre-processing.

\subsection{Stages Classification}

In terms of Stage disease, we were challenged with lack of large-scale ROP dataset with each case, especially for Stages 4, and 5. As noted previously, Stage 4 or Stage 5 data was not available, we  we instead focused on Stages 0-3. Our approach differed from other ROP papers \cite{wang2018automated} which combined various conditions e.g. Stage, Zone and Plus. Instead, using previous findings of Mulay et al., \cite{mulay2019early}, Ding et al., \cite{ding2020retinopathy} and Tong et al., \cite{tong2020automated}, this research focused on the impact image pre-processing has on ROP classifiers for Stages such that the presence of demarcation line was picked up. Our approach was to understand the problem of fundus reflection and then tuning relevant methods for ROP. 

Using the seven (7) pre-processing datasets including unprocessed baseline ROP dataset, the results are highlighted in bold in Tables ~\ref{table:TabledatasetsStageResnet50}, ~\ref{table:TabledatasetsStageinceptionresv2} for Resnet50 and InceptionResv2 CNNs respectively.  Using sensitivity and sensitivity (recall) as a primary measures, in ResNet50, PCAr performed best of all image pre-processing methods with sensitivity/specificity of 56.89\%/93.44\%. PCAr-CLAHE was second best at 53.44\%/91.72\%. In comparison, CLAHE and unprocessed images had the lowest value of 32.76\%/98.96\%, and 39.83\%/94.83\%. With InceptionResv2, we saw DPFRr-CLAHE obtain the highest sensitivity/specificity of 72.41\%/96.21\%. Precision was 79.25\%. DPFRr had the second best outcome with sensitivity/specificity of 63.79\%/94.48\%. The lowest results were from unprocessed fundus images with 39.65\%/95.86\%. The results demonstrated a pattern of improvements in ROP Stage detection when using our newer improved methods in comparison to unprocessed or standard traditional techniques. In the newer improved restoration methods, the demarcation line is more visible and additional noise relating to colour is suppressed especially when further processed with CLAHE. To illustrate this case, when we compare CLAHE with DPFRr, in either ResNet50 or InceptionResv2, we see sensitivity improve. When DPFRr-CLAHE which is DPFRr with CLAHE post processing, the sensitivity is much further improved. This is primarily because the colour factor in DPFRr is suppressed with demarcation line being better visible after processing it further with 3 channel CLAHE process.

\begin{table}[!ht]
\centering
\scalebox{0.5}{
\resizebox{\textwidth}{!}{%
\begin{tabular}{l|llllll|}
\cline{2-7}
 &
  \multicolumn{6}{c|}{\cellcolor[HTML]{DDEBF7}\textbf{ResNet50 - Stages}} \\ \hline
\rowcolor[HTML]{E7E6E6} 
\multicolumn{1}{|c|}{\cellcolor[HTML]{E7E6E6}\textbf{Methods}} &
  \multicolumn{1}{c|}{\cellcolor[HTML]{E7E6E6}\textbf{Sensitivity}} &
  \multicolumn{1}{c|}{\cellcolor[HTML]{E7E6E6}\textbf{Specificity}} &
  \multicolumn{1}{c|}{\cellcolor[HTML]{E7E6E6}\textbf{Precision}} &
  \multicolumn{1}{c|}{\cellcolor[HTML]{E7E6E6}\textbf{F1}} &
  \multicolumn{1}{c|}{\cellcolor[HTML]{E7E6E6}\textbf{Kappa}} &
  \multicolumn{1}{c|}{\cellcolor[HTML]{E7E6E6}\textbf{Accuracy}} \\ \hline
\rowcolor[HTML]{DDEBF7} 
\multicolumn{1}{|l|}{\cellcolor[HTML]{DDEBF7}Base} &
  \multicolumn{1}{l|}{\cellcolor[HTML]{DDEBF7}0.3965} &
  \multicolumn{1}{l|}{\cellcolor[HTML]{DDEBF7}0.9483} &
  \multicolumn{1}{l|}{\cellcolor[HTML]{DDEBF7}0.6053} &
  \multicolumn{1}{l|}{\cellcolor[HTML]{DDEBF7}0.4792} &
  \multicolumn{1}{l|}{\cellcolor[HTML]{DDEBF7}0.3999} &
  0.8563 \\ \hline
\multicolumn{1}{|l|}{Gray} &
  \multicolumn{1}{l|}{0.4483} &
  \multicolumn{1}{l|}{0.0448} &
  \multicolumn{1}{l|}{0.6190} &
  \multicolumn{1}{l|}{0.5200} &
  \multicolumn{1}{l|}{0.4119} &
  0.8621 \\ \hline
\rowcolor[HTML]{DDEBF7} 
\multicolumn{1}{|l|}{\cellcolor[HTML]{DDEBF7}CLAHE} &
  \multicolumn{1}{l|}{\cellcolor[HTML]{DDEBF7}0.3276} &
  \multicolumn{1}{l|}{\cellcolor[HTML]{DDEBF7}0.9896} &
  \multicolumn{1}{l|}{\cellcolor[HTML]{DDEBF7}0.8636} &
  \multicolumn{1}{l|}{\cellcolor[HTML]{DDEBF7}0.4750} &
  \multicolumn{1}{l|}{\cellcolor[HTML]{DDEBF7}0.422} &
  0.8791 \\ \hline
\multicolumn{1}{|l|}{CGH} &
  \multicolumn{1}{l|}{0.3448} &
  \multicolumn{1}{l|}{0.9483} &
  \multicolumn{1}{l|}{0.5714} &
  \multicolumn{1}{l|}{0.4301} &
  \multicolumn{1}{l|}{0.3483} &
  0.8477 \\ \hline
\rowcolor[HTML]{DDEBF7} 
\multicolumn{1}{|l|}{\cellcolor[HTML]{DDEBF7}DPFRr} &
  \multicolumn{1}{l|}{\cellcolor[HTML]{DDEBF7}0.4482} &
  \multicolumn{1}{l|}{\cellcolor[HTML]{DDEBF7}0.9552} &
  \multicolumn{1}{l|}{\cellcolor[HTML]{DDEBF7}0.6667} &
  \multicolumn{1}{l|}{\cellcolor[HTML]{DDEBF7}0.5361} &
  \multicolumn{1}{l|}{\cellcolor[HTML]{DDEBF7}0.4642} &
  0.8707 \\ \hline
\multicolumn{1}{|l|}{DPFRr\_CLAHE} &
  \multicolumn{1}{l|}{0.4482} &
  \multicolumn{1}{l|}{0.9217} &
  \multicolumn{1}{l|}{0.6500} &
  \multicolumn{1}{l|}{0.5306} &
  \multicolumn{1}{l|}{0.4567} &
  0.8678 \\ \hline
\rowcolor[HTML]{FFFF00} 
\multicolumn{1}{|l|}{\cellcolor[HTML]{FFFF00}\textbf{PCAr}} &
  \multicolumn{1}{l|}{\cellcolor[HTML]{FFFF00}\textbf{0.5689}} &
  \multicolumn{1}{l|}{\cellcolor[HTML]{FFFF00}\textbf{0.9344}} &
  \multicolumn{1}{l|}{\cellcolor[HTML]{FFFF00}\textbf{0.6346}} &
  \multicolumn{1}{l|}{\cellcolor[HTML]{FFFF00}\textbf{0.6000}} &
  \multicolumn{1}{l|}{\cellcolor[HTML]{FFFF00}\textbf{0.5252}} &
  \textbf{0.8736} \\ \hline
\multicolumn{1}{|l|}{PCAr\_CLAHE} &
  \multicolumn{1}{l|}{0.5344} &
  \multicolumn{1}{l|}{0.9172} &
  \multicolumn{1}{l|}{0.5636} &
  \multicolumn{1}{l|}{0.5487} &
  \multicolumn{1}{l|}{0.4617} &
  0.8534 \\ \hline
\end{tabular}%
}
}
\caption{Resnet50 results for Stages classification for each pre-processing method }
\label{table:TabledatasetsStageResnet50}
\end{table}

\begin{table}[!ht]
\centering
\scalebox{0.5}{
\resizebox{\textwidth}{!}{%
\begin{tabular}{l|llllll|}
\cline{2-7}
 &
  \multicolumn{6}{c|}{\cellcolor[HTML]{DDEBF7}\textbf{InceptionResv2 - Stages}} \\ \hline
\rowcolor[HTML]{E7E6E6} 
\multicolumn{1}{|c|}{\cellcolor[HTML]{E7E6E6}\textbf{Methods}} &
  \multicolumn{1}{c|}{\cellcolor[HTML]{E7E6E6}\textbf{Sensitivity}} &
  \multicolumn{1}{c|}{\cellcolor[HTML]{E7E6E6}\textbf{Specificity}} &
  \multicolumn{1}{c|}{\cellcolor[HTML]{E7E6E6}\textbf{Precision}} &
  \multicolumn{1}{c|}{\cellcolor[HTML]{E7E6E6}\textbf{F1}} &
  \multicolumn{1}{c|}{\cellcolor[HTML]{E7E6E6}\textbf{Kappa}} &
  \multicolumn{1}{c|}{\cellcolor[HTML]{E7E6E6}\textbf{Accuracy}} \\ \hline
\rowcolor[HTML]{DDEBF7} 
\multicolumn{1}{|l|}{\cellcolor[HTML]{DDEBF7}Base} &
  \multicolumn{1}{l|}{\cellcolor[HTML]{DDEBF7}0.3965} &
  \multicolumn{1}{l|}{\cellcolor[HTML]{DDEBF7}0.9586} &
  \multicolumn{1}{l|}{\cellcolor[HTML]{DDEBF7}0.6571} &
  \multicolumn{1}{l|}{\cellcolor[HTML]{DDEBF7}0.4946} &
  \multicolumn{1}{l|}{\cellcolor[HTML]{DDEBF7}0.4221} &
  0.8649 \\ \hline
\multicolumn{1}{|l|}{Gray} &
  \multicolumn{1}{l|}{0.5000} &
  \multicolumn{1}{l|}{0.9241} &
  \multicolumn{1}{l|}{0.6667} &
  \multicolumn{1}{l|}{0.5321} &
  \multicolumn{1}{l|}{0.4456} &
  0.8534 \\ \hline
\rowcolor[HTML]{DDEBF7} 
\multicolumn{1}{|l|}{\cellcolor[HTML]{DDEBF7}CLAHE} &
  \multicolumn{1}{l|}{\cellcolor[HTML]{DDEBF7}0.6034} &
  \multicolumn{1}{l|}{\cellcolor[HTML]{DDEBF7}0.9276} &
  \multicolumn{1}{l|}{\cellcolor[HTML]{DDEBF7}0.6250} &
  \multicolumn{1}{l|}{\cellcolor[HTML]{DDEBF7}0.6140} &
  \multicolumn{1}{l|}{\cellcolor[HTML]{DDEBF7}0.5386} &
  0.8736 \\ \hline
\multicolumn{1}{|l|}{CGH} &
  \multicolumn{1}{l|}{0.4828} &
  \multicolumn{1}{l|}{0.9207} &
  \multicolumn{1}{l|}{0.5490} &
  \multicolumn{1}{l|}{0.5138} &
  \multicolumn{1}{l|}{0.4239} &
  0.8477 \\ \hline
\rowcolor[HTML]{DDEBF7} 
\multicolumn{1}{|l|}{\cellcolor[HTML]{DDEBF7}DPFRr} &
  \multicolumn{1}{l|}{\cellcolor[HTML]{DDEBF7}0.6379} &
  \multicolumn{1}{l|}{\cellcolor[HTML]{DDEBF7}0.9448} &
  \multicolumn{1}{l|}{\cellcolor[HTML]{DDEBF7}0.6981} &
  \multicolumn{1}{l|}{\cellcolor[HTML]{DDEBF7}0.6667} &
  \multicolumn{1}{l|}{\cellcolor[HTML]{DDEBF7}0.6036} &
  0.8936 \\ \hline
\rowcolor[HTML]{FFFF00} 
\multicolumn{1}{|l|}{\cellcolor[HTML]{FFFF00}\textbf{DPFRr\_CLAHE}} &
  \multicolumn{1}{l|}{\cellcolor[HTML]{FFFF00}\textbf{0.7241}} &
  \multicolumn{1}{l|}{\cellcolor[HTML]{FFFF00}\textbf{0.9621}} &
  \multicolumn{1}{l|}{\cellcolor[HTML]{FFFF00}\textbf{0.7925}} &
  \multicolumn{1}{l|}{\cellcolor[HTML]{FFFF00}\textbf{0.7567}} &
  \multicolumn{1}{l|}{\cellcolor[HTML]{FFFF00}\textbf{0.7107}} &
  \textbf{0.9224} \\ \hline
\rowcolor[HTML]{DDEBF7} 
\multicolumn{1}{|l|}{\cellcolor[HTML]{DDEBF7}PCAr} &
  \multicolumn{1}{l|}{\cellcolor[HTML]{DDEBF7}0.5344} &
  \multicolumn{1}{l|}{\cellcolor[HTML]{DDEBF7}0.9207} &
  \multicolumn{1}{l|}{\cellcolor[HTML]{DDEBF7}0.5740} &
  \multicolumn{1}{l|}{\cellcolor[HTML]{DDEBF7}0.5536} &
  \multicolumn{1}{l|}{\cellcolor[HTML]{DDEBF7}0.468} &
  0.8563 \\ \hline
\multicolumn{1}{|l|}{PCAr\_CLAHE} &
  \multicolumn{1}{l|}{0.5172} &
  \multicolumn{1}{l|}{0.9207} &
  \multicolumn{1}{l|}{0.5660} &
  \multicolumn{1}{l|}{0.5405} &
  \multicolumn{1}{l|}{0.4535} &
  0.8535 \\ \hline
\end{tabular}%
}
}
\caption{InceptionResv2 results for Stage classification for each pre-processing method }
\label{table:TabledatasetsStageinceptionresv2}
\end{table}

\begin{table}[!ht]
\resizebox{\columnwidth}{!}{%
\begin{tabular}{l|lllll|l|}
\cline{2-7}
\textbf{} &
  \multicolumn{5}{l|}{\cellcolor[HTML]{DDEBF7}\textbf{Performance   of DPRFr-InceptionResv2 Architecture - Stages}} &
  \cellcolor[HTML]{DDEBF7}\textbf{} \\ \hline
\rowcolor[HTML]{E7E6E6} 
\multicolumn{1}{|l|}{\cellcolor[HTML]{E7E6E6}\textbf{Type}} &
  \multicolumn{1}{l|}{\cellcolor[HTML]{E7E6E6}\textbf{Features}} &
  \multicolumn{1}{l|}{\cellcolor[HTML]{E7E6E6}\textbf{Sensitivity}} &
  \multicolumn{1}{l|}{\cellcolor[HTML]{E7E6E6}\textbf{Specificity}} &
  \multicolumn{1}{l|}{\cellcolor[HTML]{E7E6E6}\textbf{Precision}} &
  \textbf{F1 Score} &
  \textbf{Accuracy} \\ \hline
\multicolumn{1}{|l|}{\cellcolor[HTML]{DDEBF7}\textbf{Stage 0}} &
  \multicolumn{1}{l|}{No Stage disease} &
  \multicolumn{1}{l|}{0.6300} &
  \multicolumn{1}{l|}{0.8800} &
  \multicolumn{1}{l|}{0.6700} &
  0.6500 &
  0.8103 \\ \hline
\multicolumn{1}{|l|}{\cellcolor[HTML]{DDEBF7}\textbf{Stage 1}} &
  \multicolumn{1}{l|}{Presence of demarcation line} &
  \multicolumn{1}{l|}{0.6000} &
  \multicolumn{1}{l|}{0.9400} &
  \multicolumn{1}{l|}{0.6700} &
  0.6300 &
  0.8793 \\ \hline
\multicolumn{1}{|l|}{\cellcolor[HTML]{DDEBF7}\textbf{Stage 2}} &
  \multicolumn{1}{l|}{Presence of ridge} &
  \multicolumn{1}{l|}{0.8300} &
  \multicolumn{1}{l|}{0.8800} &
  \multicolumn{1}{l|}{0.7500} &
  0.7900 &
  0.8621 \\ \hline
\multicolumn{1}{|l|}{\cellcolor[HTML]{DDEBF7}\textbf{Stage 3}} &
  \multicolumn{1}{l|}{Ridge with additional features} &
  \multicolumn{1}{l|}{0.8600} &
  \multicolumn{1}{l|}{0.9500} &
  \multicolumn{1}{l|}{0.8600} &
  0.8600 &
  0.9310 \\ \hline
\end{tabular}%
}
\caption{Stages Classification breakdown for DPFRr-CLAHE-InceptionResv2}
\label{table:TabledatasetsB1Stages}
\end{table}

\begin{table}[!ht]
\begin{tabular}{c|llll|}
\cline{2-5}
\multicolumn{1}{l|}{} &
  \multicolumn{4}{c|}{\cellcolor[HTML]{DAE8FC}\textbf{Confusion   Matrix}} \\ \hline
\rowcolor[HTML]{DAE8FC} 
\multicolumn{1}{|c|}{\cellcolor[HTML]{DAE8FC}\textbf{STAGES}} &
  \multicolumn{1}{c|}{\cellcolor[HTML]{DAE8FC}\textbf{Stage 0}} &
  \multicolumn{1}{c|}{\cellcolor[HTML]{DAE8FC}\textbf{Stage 1}} &
  \multicolumn{1}{c|}{\cellcolor[HTML]{DAE8FC}\textbf{Stage 2}} &
  \multicolumn{1}{c|}{\cellcolor[HTML]{DAE8FC}\textbf{Stage 3}} \\ \hline
\multicolumn{1}{|c|}{\cellcolor[HTML]{DAE8FC}\textbf{Stage 0}} &
  \multicolumn{1}{l|}{\cellcolor[HTML]{DDEBF7}10} &
  \multicolumn{1}{l|}{2} &
  \multicolumn{1}{l|}{2} &
  1 \\ \hline
\multicolumn{1}{|c|}{\cellcolor[HTML]{DAE8FC}\textbf{Stage 1}} &
  \multicolumn{1}{l|}{2} &
  \multicolumn{1}{l|}{\cellcolor[HTML]{DDEBF7}6} &
  \multicolumn{1}{l|}{1} &
  0 \\ \hline
\multicolumn{1}{|c|}{\cellcolor[HTML]{DAE8FC}\textbf{Stage 2}} &
  \multicolumn{1}{l|}{3} &
  \multicolumn{1}{l|}{1} &
  \multicolumn{1}{l|}{\cellcolor[HTML]{DDEBF7}15} &
  1 \\ \hline
\multicolumn{1}{|c|}{\cellcolor[HTML]{DAE8FC}\textbf{Stage 3}} &
  \multicolumn{1}{l|}{1} &
  \multicolumn{1}{l|}{1} &
  \multicolumn{1}{l|}{0} &
  \cellcolor[HTML]{DDEBF7}12 \\ \hline
\end{tabular}
\caption{Confusion Matrix for Stages result for DPFRr-CLAHE-InceptionResv2}
\label{table:TabledatasetsC1Stages}
\end{table}

Amongst all methods/CNN combination, we found DPFRr-CLAHE using InceptionResv2 CNN gave best sensitivity/specificity of 72.41\%/96.21\%. Its detailed breakdown is shown in Table~\ref{table:TabledatasetsB1Stages} with its confusion matrix in Table~\ref{table:TabledatasetsC1Stages}.  DPFRr-CLAHE/InceptionResv2 results breakdown showed that it provided a high sensitivity/specificity/precision for Stage 0 (63\%/67\%/88\%) followed by Stage 1 (60\%/67\%/94\%), Stage 2 (83\%/75\%/88\%) and Stage 3 (86\%/86\%/95\%) respectively.  This can be explained with the following insights. Stage 1 classification was still challenged primarily due to the very fine grainy output which, though reduced significantly did cause challenges in detecting the faintest of demarcation line. Improving Stage 1 detection is part of future work of improvement. Stage 2 and Stage 3 show significant improvements in classification primarily because the demarcation line is now very visible. This was as of the point of writing, the only research where Stage classification for ROP Retcam, had been performed using a restoration based image pre-processing.

\subsection{Zones Classification}

\begin{table}[!ht]
\centering
\scalebox{0.5}{
\resizebox{\textwidth}{!}{%
\begin{tabular}{l|llllll|}
\cline{2-7}
 &
  \multicolumn{6}{c|}{\cellcolor[HTML]{DDEBF7}\textbf{ResNet50   - Zones}} \\ \hline
\rowcolor[HTML]{E7E6E6} 
\multicolumn{1}{|c|}{\cellcolor[HTML]{E7E6E6}\textbf{Methods}} &
  \multicolumn{1}{c|}{\cellcolor[HTML]{E7E6E6}\textbf{Sensitivity}} &
  \multicolumn{1}{c|}{\cellcolor[HTML]{E7E6E6}\textbf{Specificity}} &
  \multicolumn{1}{c|}{\cellcolor[HTML]{E7E6E6}\textbf{Precision}} &
  \multicolumn{1}{c|}{\cellcolor[HTML]{E7E6E6}\textbf{F1}} &
  \multicolumn{1}{c|}{\cellcolor[HTML]{E7E6E6}\textbf{Kappa}} &
  \multicolumn{1}{c|}{\cellcolor[HTML]{E7E6E6}\textbf{Accuracy}} \\ \hline
\rowcolor[HTML]{DDEBF7} 
\multicolumn{1}{|l|}{\cellcolor[HTML]{DDEBF7}Base} &
  \multicolumn{1}{l|}{\cellcolor[HTML]{DDEBF7}0.7805} &
  \multicolumn{1}{l|}{\cellcolor[HTML]{DDEBF7}0.9146} &
  \multicolumn{1}{l|}{\cellcolor[HTML]{DDEBF7}0.8205} &
  \multicolumn{1}{l|}{\cellcolor[HTML]{DDEBF7}0.8000} &
  \multicolumn{1}{l|}{\cellcolor[HTML]{DDEBF7}0.7037} &
  0.8699 \\ \hline
\multicolumn{1}{|l|}{Gray} &
  \multicolumn{1}{l|}{0.7073} &
  \multicolumn{1}{l|}{0.8536} &
  \multicolumn{1}{l|}{0.7073} &
  \multicolumn{1}{l|}{0.7073} &
  \multicolumn{1}{l|}{0.5610} &
  0.8699 \\ \hline
\rowcolor[HTML]{DDEBF7} 
\multicolumn{1}{|l|}{\cellcolor[HTML]{DDEBF7}CLAHE} &
  \multicolumn{1}{l|}{\cellcolor[HTML]{DDEBF7}0.7805} &
  \multicolumn{1}{l|}{\cellcolor[HTML]{DDEBF7}0.9024} &
  \multicolumn{1}{l|}{\cellcolor[HTML]{DDEBF7}0.8000} &
  \multicolumn{1}{l|}{\cellcolor[HTML]{DDEBF7}0.7912} &
  \multicolumn{1}{l|}{\cellcolor[HTML]{DDEBF7}0.6871} &
  0.8618 \\ \hline
\multicolumn{1}{|l|}{CGH} &
  \multicolumn{1}{l|}{0.7317} &
  \multicolumn{1}{l|}{0.8902} &
  \multicolumn{1}{l|}{0.7692} &
  \multicolumn{1}{l|}{0.8373} &
  \multicolumn{1}{l|}{0.6296} &
  0.8374 \\ \hline
\rowcolor[HTML]{DDEBF7} 
\multicolumn{1}{|l|}{\cellcolor[HTML]{DDEBF7}DPFRr} &
  \multicolumn{1}{l|}{\cellcolor[HTML]{DDEBF7}0.7317} &
  \multicolumn{1}{l|}{\cellcolor[HTML]{DDEBF7}0.8780} &
  \multicolumn{1}{l|}{\cellcolor[HTML]{DDEBF7}0.7500} &
  \multicolumn{1}{l|}{\cellcolor[HTML]{DDEBF7}0.7407} &
  \multicolumn{1}{l|}{\cellcolor[HTML]{DDEBF7}0.6134} &
  0.8293 \\ \hline
\rowcolor[HTML]{FFFF00} 
\multicolumn{1}{|l|}{\cellcolor[HTML]{FFFF00}\textbf{DPFRr\_CLAHE}} &
  \multicolumn{1}{l|}{\cellcolor[HTML]{FFFF00}\textbf{0.8537}} &
  \multicolumn{1}{l|}{\cellcolor[HTML]{FFFF00}\textbf{0.9268}} &
  \multicolumn{1}{l|}{\cellcolor[HTML]{FFFF00}\textbf{0.8536}} &
  \multicolumn{1}{l|}{\cellcolor[HTML]{FFFF00}\textbf{0.8536}} &
  \multicolumn{1}{l|}{\cellcolor[HTML]{FFFF00}\textbf{0.7804}} &
  \textbf{0.9024} \\ \hline
\rowcolor[HTML]{DDEBF7} 
\multicolumn{1}{|l|}{\cellcolor[HTML]{DDEBF7}PCAr} &
  \multicolumn{1}{l|}{\cellcolor[HTML]{DDEBF7}0.7561} &
  \multicolumn{1}{l|}{\cellcolor[HTML]{DDEBF7}0.8781} &
  \multicolumn{1}{l|}{\cellcolor[HTML]{DDEBF7}0.7561} &
  \multicolumn{1}{l|}{\cellcolor[HTML]{DDEBF7}0.7561} &
  \multicolumn{1}{l|}{\cellcolor[HTML]{DDEBF7}0.6341} &
  0.8374 \\ \hline
\multicolumn{1}{|l|}{PCAr\_CLAHE} &
  \multicolumn{1}{l|}{0.7073} &
  \multicolumn{1}{l|}{0.9024} &
  \multicolumn{1}{l|}{0.7838} &
  \multicolumn{1}{l|}{0.8374} &
  \multicolumn{1}{l|}{0.6250} &
  0.8374 \\ \hline
\end{tabular}%
}
}
\caption{ResNet50 results for Zones classification for each pre-processing method }
\label{table:TabledatasetsZoneResnet50}
\end{table}

\begin{table}[!ht]
\centering
\scalebox{0.5}{
\resizebox{\textwidth}{!}{%
\begin{tabular}{l|llllll|}
\cline{2-7}
 &
  \multicolumn{6}{c|}{\cellcolor[HTML]{DDEBF7}\textbf{InceptionResv2 -Zones}} \\ \hline
\rowcolor[HTML]{E7E6E6} 
\multicolumn{1}{|c|}{\cellcolor[HTML]{E7E6E6}\textbf{Methods}} &
  \multicolumn{1}{c|}{\cellcolor[HTML]{E7E6E6}\textbf{Sensitivity}} &
  \multicolumn{1}{c|}{\cellcolor[HTML]{E7E6E6}\textbf{Specificity}} &
  \multicolumn{1}{c|}{\cellcolor[HTML]{E7E6E6}\textbf{Precision}} &
  \multicolumn{1}{c|}{\cellcolor[HTML]{E7E6E6}\textbf{F1}} &
  \multicolumn{1}{c|}{\cellcolor[HTML]{E7E6E6}\textbf{Kappa}} &
  \multicolumn{1}{c|}{\cellcolor[HTML]{E7E6E6}\textbf{Accuracy}} \\ \hline
\rowcolor[HTML]{DDEBF7} 
\multicolumn{1}{|l|}{\cellcolor[HTML]{DDEBF7}Base} &
  \multicolumn{1}{l|}{\cellcolor[HTML]{DDEBF7}0.5854} &
  \multicolumn{1}{l|}{\cellcolor[HTML]{DDEBF7}0.8902} &
  \multicolumn{1}{l|}{\cellcolor[HTML]{DDEBF7}0.7273} &
  \multicolumn{1}{l|}{\cellcolor[HTML]{DDEBF7}0.6486} &
  \multicolumn{1}{l|}{\cellcolor[HTML]{DDEBF7}0.5000} &
  0.7886 \\ \hline
\multicolumn{1}{|l|}{Gray} &
  \multicolumn{1}{l|}{0.4146} &
  \multicolumn{1}{l|}{0.8536} &
  \multicolumn{1}{l|}{0.5862} &
  \multicolumn{1}{l|}{0.4857} &
  \multicolumn{1}{l|}{0.2894} &
  0.7073 \\ \hline
\rowcolor[HTML]{DDEBF7} 
\multicolumn{1}{|l|}{\cellcolor[HTML]{DDEBF7}CLAHE} &
  \multicolumn{1}{l|}{\cellcolor[HTML]{DDEBF7}0.5609} &
  \multicolumn{1}{l|}{\cellcolor[HTML]{DDEBF7}0.8170} &
  \multicolumn{1}{l|}{\cellcolor[HTML]{DDEBF7}0.6052} &
  \multicolumn{1}{l|}{\cellcolor[HTML]{DDEBF7}0.5822} &
  \multicolumn{1}{l|}{\cellcolor[HTML]{DDEBF7}0.3850} &
  0.7317 \\ \hline
\multicolumn{1}{|l|}{CGH} &
  \multicolumn{1}{l|}{0.5854} &
  \multicolumn{1}{l|}{0.8537} &
  \multicolumn{1}{l|}{0.6667} &
  \multicolumn{1}{l|}{0.6234} &
  \multicolumn{1}{l|}{0.4528} &
  0.7642 \\ \hline
\rowcolor[HTML]{DDEBF7} 
\multicolumn{1}{|l|}{\cellcolor[HTML]{DDEBF7}DPFRr} &
  \multicolumn{1}{l|}{\cellcolor[HTML]{DDEBF7}0.5121} &
  \multicolumn{1}{l|}{\cellcolor[HTML]{DDEBF7}0.8283} &
  \multicolumn{1}{l|}{\cellcolor[HTML]{DDEBF7}0.6000} &
  \multicolumn{1}{l|}{\cellcolor[HTML]{DDEBF7}0.5526} &
  \multicolumn{1}{l|}{\cellcolor[HTML]{DDEBF7}0.4528} &
  0.7236 \\ \hline
\multicolumn{1}{|l|}{DPFRr\_CLAHE} &
  \multicolumn{1}{l|}{0.5122} &
  \multicolumn{1}{l|}{0.8537} &
  \multicolumn{1}{l|}{0.6363} &
  \multicolumn{1}{l|}{0.5676} &
  \multicolumn{1}{l|}{0.3846} &
  0.7398 \\ \hline
\rowcolor[HTML]{DDEBF7} 
\multicolumn{1}{|l|}{\cellcolor[HTML]{DDEBF7}PCAr} &
  \multicolumn{1}{l|}{\cellcolor[HTML]{DDEBF7}0.5122} &
  \multicolumn{1}{l|}{\cellcolor[HTML]{DDEBF7}0.8049} &
  \multicolumn{1}{l|}{\cellcolor[HTML]{DDEBF7}0.5676} &
  \multicolumn{1}{l|}{\cellcolor[HTML]{DDEBF7}0.5385} &
  \multicolumn{1}{l|}{\cellcolor[HTML]{DDEBF7}0.3250} &
  0.7073 \\ \hline
\rowcolor[HTML]{FFFF00} 
\multicolumn{1}{|l|}{\cellcolor[HTML]{FFFF00}\textbf{PCAr\_CLAHE}} &
  \multicolumn{1}{l|}{\cellcolor[HTML]{FFFF00}\textbf{0.6097}} &
  \multicolumn{1}{l|}{\cellcolor[HTML]{FFFF00}\textbf{0.8095}} &
  \multicolumn{1}{l|}{\cellcolor[HTML]{FFFF00}\textbf{0.6098}} &
  \multicolumn{1}{l|}{\cellcolor[HTML]{FFFF00}\textbf{0.6098}} &
  \multicolumn{1}{l|}{\cellcolor[HTML]{FFFF00}\textbf{0.4444}} &
  \textbf{0.7440} \\ \hline
\end{tabular}%
}
}
\caption{InceptionResv2 results for Zones classification for each pre-processing method }
\label{table:TabledatasetsZoneInceptionResv2}
\end{table}


\begin{table}[!ht]
\resizebox{\columnwidth}{!}{%
\begin{tabular}{l|llllll|}
\cline{2-7}
\textbf{} &
  \multicolumn{6}{c|}{\cellcolor[HTML]{DDEBF7}\textbf{Performance   of DPRFr-InceptionResv2 Architecture - Zones}} \\ \hline
\rowcolor[HTML]{E7E6E6} 
\multicolumn{1}{|l|}{\cellcolor[HTML]{E7E6E6}\textbf{Type}} &
  \multicolumn{1}{l|}{\cellcolor[HTML]{E7E6E6}\textbf{Features}} &
  \multicolumn{1}{l|}{\cellcolor[HTML]{E7E6E6}\textbf{Sensitivity}} &
  \multicolumn{1}{l|}{\cellcolor[HTML]{E7E6E6}\textbf{Specificity}} &
  \multicolumn{1}{l|}{\cellcolor[HTML]{E7E6E6}\textbf{Precision}} &
  \multicolumn{1}{l|}{\cellcolor[HTML]{E7E6E6}\textbf{F1 Score}} &
  \textbf{Accuracy} \\ \hline
\multicolumn{1}{|l|}{\cellcolor[HTML]{DDEBF7}\textbf{Zone I}} &
  \multicolumn{1}{l|}{Presence of disease in Zone I} &
  \multicolumn{1}{l|}{1.0000} &
  \multicolumn{1}{l|}{0.9744} &
  \multicolumn{1}{l|}{0.6700} &
  \multicolumn{1}{l|}{0.8000} &
  0.9756 \\ \hline
\multicolumn{1}{|l|}{\cellcolor[HTML]{DDEBF7}\textbf{Zone II}} &
  \multicolumn{1}{l|}{Presence of disease in Zone II} &
  \multicolumn{1}{l|}{0.8462} &
  \multicolumn{1}{l|}{1.0000} &
  \multicolumn{1}{l|}{1.0000} &
  \multicolumn{1}{l|}{0.9167} &
  0.8537 \\ \hline
\multicolumn{1}{|l|}{\cellcolor[HTML]{DDEBF7}\textbf{Zone III}} &
  \multicolumn{1}{l|}{Presence of disease in Zone III} &
  \multicolumn{1}{l|}{0.0000} &
  \multicolumn{1}{l|}{0.8780} &
  \multicolumn{1}{l|}{0.0000} &
  \multicolumn{1}{l|}{0.0000} &
  0.8780 \\ \hline
\end{tabular}%
}
\caption{Zones Classification breakdown for DPFRr-CLAHE ResNet50}
\label{table:TabledatasetsB1Zones}
\end{table}

\begin{table}[!ht]
\begin{tabular}{|
>{\columncolor[HTML]{DAE8FC}}l lll|}
\hline
\multicolumn{4}{|c|}{\cellcolor[HTML]{DAE8FC}\textbf{Confusion   Matrix}}                                                                                                                                                                  \\ \hline
\multicolumn{1}{|l|}{\cellcolor[HTML]{DAE8FC}\textbf{Zones}}    & \multicolumn{1}{l|}{\cellcolor[HTML]{DAE8FC}\textbf{Zone I}} & \multicolumn{1}{l|}{\cellcolor[HTML]{DAE8FC}\textbf{Zone II}} & \cellcolor[HTML]{DAE8FC}\textbf{Zone III} \\ \hline
\multicolumn{1}{|l|}{\cellcolor[HTML]{DAE8FC}\textbf{Zone I}}   & \multicolumn{1}{l|}{\cellcolor[HTML]{ECF4FF}2}               & \multicolumn{1}{l|}{1}                                        & 0                                         \\ \hline
\multicolumn{1}{|l|}{\cellcolor[HTML]{DAE8FC}\textbf{Zone II}}  & \multicolumn{1}{l|}{0}                                       & \multicolumn{1}{l|}{\cellcolor[HTML]{ECF4FF}33}               & 0                                         \\ \hline
\multicolumn{1}{|l|}{\cellcolor[HTML]{DAE8FC}\textbf{Zone III}} & \multicolumn{1}{l|}{0}                                       & \multicolumn{1}{l|}{5}                                        & \cellcolor[HTML]{ECF4FF}0                 \\ \hline
\end{tabular}
\caption{Confusion Matrix for Zones Classification}
\label{table:TabledatasetsC1Zones}
\end{table}

In our results, we observed that DPRFr-CLAHE using ResNet50 worked well here. It provided high sensitivity/specificity of 85.37\%/92.68\%.  Precision was 85.36\%. Further breakdown analysis of DPFRr/ResNet50 results identified Zone I/II with sensitivity/specificity/precision and overall accuracy. In terms of Zone I detection, it showed sensitivity/specificity/precision of 100\%/97\%/67\% (accuracy of 93\%), and Zone II with 84\%/100\%/85\% (accuracy of 85.37\%) respectively. Zone III was challenged due to lack of sufficient data. Zhao \cite{zhao2019deep} in comparison only obtained 91\% for Zone I. He did not attempt Zone II, or Zone III. This work was, as far as we are aware, where Zone I and II had been shown to be classified successfully using deep learning. 

Overall, there was improvement noted for PCAr when combined with CLAHE. Similarly same effects were noted for DPFRr-CLAHE as well.

\section{Discussion}
\label{sec:Discussion}

In lieu of the results obtained and comparative analysis with other related works, we demonstrated that there were significant improvements as a result of improved image pre-processing that improved accuracy for Plus disease. It was very significantly improved for Stages and to some extent Zones.

For Plus disease classification, our improved novel ROP specific processes (PCAr, PCAr-CLAHE, DPFRr, DPFRr-CLAHE) showed improvements over other methods in classification when used with the transfer learning based CNNs. Both DPFRr and DPFRr-CLAHE showed improvements primarily due to the removal of the source of reflection as well as the reduction on the choroid vessels. This allowed primary blood vessels where tortuosity was present to be more clearly visible. Applying CLAHE to the DPFRr reduced the colour factor further.  The comparative analysis as noted in the table~\ref{table:TabledatasetsC2Plus} below suggested that these new hybrids showed better results when compared to other research papers where R-CNN was used as well. Analysing our results for  DPFRr-CLAHE/InceptionResv2 CNN results, a high precision, sensitivity, specificity for NoPlus (100\%/97\%/100\%) and Plus (85\%/100\%/97\%/65\%) was obtained. This was similar to results obtained by iROP-DL \cite{brown2018automated}, and better than Vinekar et al., \cite{vinekar2021development}. 

\begin{table}[!ht]
\scalebox{1} {
\resizebox{\columnwidth}{!}{%
\begin{tabular}{l|llll|}
\cline{2-5}
                                  & \multicolumn{4}{c|}{\cellcolor[HTML]{DDEBF7}\textbf{Results Comparison}}                 \\ \hline
\rowcolor[HTML]{E7E6E6} 
\multicolumn{1}{|c|}{\cellcolor[HTML]{E7E6E6}\textbf{Measure}} &
  \multicolumn{1}{c|}{\cellcolor[HTML]{E7E6E6}\textbf{McROP}} &
  \multicolumn{1}{c|}{\cellcolor[HTML]{E7E6E6}\textbf{Brown et al.,\cite{brown2018automated}}} &
  \multicolumn{1}{c|}{\cellcolor[HTML]{E7E6E6}\textbf{Tong et al.,\cite{tong2020automated} }} &
  \multicolumn{1}{c|}{\cellcolor[HTML]{E7E6E6}\textbf{Vinekar et al.,\cite{vinekar2021development} }} \\ \hline
\rowcolor[HTML]{DDEBF7} 
\multicolumn{1}{|l|}{\cellcolor[HTML]{DDEBF7}Sensitivity} &
  \multicolumn{1}{l|}{\cellcolor[HTML]{DDEBF7}0.98} &
  \multicolumn{1}{l|}{\cellcolor[HTML]{DDEBF7}0.93} &
  \multicolumn{1}{l|}{\cellcolor[HTML]{DDEBF7}0.71} &
  0.95 \\ \hline
\multicolumn{1}{|l|}{Specificity} & \multicolumn{1}{l|}{1.00} & \multicolumn{1}{l|}{0.94} & \multicolumn{1}{l|}{0.91} & 1.00 \\ \hline
\end{tabular}
}
}
\caption{Comparison of McROP DPFRr-CLAHE InceptionResv2 for Plus Diseases}
\label{table:TabledatasetsC2Plus}
\end{table}

The described methods in conjunction with CLAHE did provide a very significant improvement over tradition and rival R-CNN methods by removing the source of the fundus image problem. Namely, reflection cancellation, and colour reduction to amplify the ROP features for Plus disease. We propose that R-CNN methods can still be used to further improve the classifiers as further add on methods for Plus disease.

In terms of Stages classification, it was important to understand the difference of improvements between DPFRr and DPFRr-CLAHE. The analysis of results and pre-processed images showed two critical findings emerge. First DPFRr significantly reduced reflection and improved retinal features for ROP but colorations was still present which challenged the classifiers.  Once the 3 channel based CLAHE applied, there was a significant further reduction in colours which made the demarcation line more pronounced. 3 channel based CLAHE application to DPFRr provided a better ROP Stage featured image allowing the classifier to better determine the stage classification. This was critical when it comes to Stages 2-3 in comparison to two comparative papers without leveraging R-CNN. This combination provided a very significant improvement over tradition and rival R-CNN methods by removing the source of the ROP Retcam image problem. As noted, addition of R-CNN further to this work should improve the results further.

\begin{table}[!ht]
\centering
\scalebox{0.5}{
\resizebox{\textwidth}{!}{%
\begin{tabular}{
>{\columncolor[HTML]{FFFFC7}}c |
>{\columncolor[HTML]{DAE8FC}}c 
>{\columncolor[HTML]{DAE8FC}}c 
>{\columncolor[HTML]{DAE8FC}}c 
>{\columncolor[HTML]{DAE8FC}}c |
>{\columncolor[HTML]{9AFF99}}c 
>{\columncolor[HTML]{9AFF99}}c 
>{\columncolor[HTML]{9AFF99}}c 
>{\columncolor[HTML]{9AFF99}}c |}
\cline{2-9}
\multicolumn{1}{l|}{\cellcolor[HTML]{FFFFFF}} &
  \multicolumn{4}{c|}{\cellcolor[HTML]{DAE8FC}\textbf{McROP}} &
  \multicolumn{4}{c|}{\cellcolor[HTML]{9AFF99}\textbf{Ding et al   (Hybrid)}} \\ \hline
\multicolumn{1}{|c|}{\cellcolor[HTML]{FFFE65}\textbf{Stages}} &
  \multicolumn{1}{c|}{\cellcolor[HTML]{C0C0C0}\textbf{Precision}} &
  \multicolumn{1}{c|}{\cellcolor[HTML]{C0C0C0}\textbf{Recall}} &
  \multicolumn{1}{c|}{\cellcolor[HTML]{C0C0C0}\textbf{Specificity}} &
  \cellcolor[HTML]{C0C0C0}\textbf{F1} &
  \multicolumn{1}{c|}{\cellcolor[HTML]{C0C0C0}\textbf{Precision}} &
  \multicolumn{1}{c|}{\cellcolor[HTML]{C0C0C0}\textbf{Recall}} &
  \multicolumn{1}{c|}{\cellcolor[HTML]{C0C0C0}\textbf{Specificity}} &
  \cellcolor[HTML]{C0C0C0}\textbf{F1} \\ \hline
\multicolumn{1}{|c|}{\cellcolor[HTML]{FFFFC7}\textbf{Stage 0}} &
  \multicolumn{1}{c|}{\cellcolor[HTML]{DAE8FC}0.67} &
  \multicolumn{1}{c|}{\cellcolor[HTML]{DAE8FC}0.63} &
  \multicolumn{1}{c|}{\cellcolor[HTML]{DAE8FC}0.88} &
  0.65 &
  \multicolumn{1}{c|}{\cellcolor[HTML]{9AFF99}-} &
  \multicolumn{1}{c|}{\cellcolor[HTML]{9AFF99}-} &
  \multicolumn{1}{c|}{\cellcolor[HTML]{9AFF99}-} &
  - \\ \hline
\multicolumn{1}{|c|}{\cellcolor[HTML]{FFFFC7}\textbf{Stage 1}} &
  \multicolumn{1}{c|}{\cellcolor[HTML]{DAE8FC}0.67} &
  \multicolumn{1}{c|}{\cellcolor[HTML]{DAE8FC}0.6} &
  \multicolumn{1}{c|}{\cellcolor[HTML]{DAE8FC}0.94} &
  0.63 &
  \multicolumn{1}{c|}{\cellcolor[HTML]{9AFF99}0.78} &
  \multicolumn{1}{c|}{\cellcolor[HTML]{9AFF99}0.77} &
  \multicolumn{1}{c|}{\cellcolor[HTML]{9AFF99}-} &
  0.78 \\ \hline
\multicolumn{1}{|c|}{\cellcolor[HTML]{FFFFC7}\textbf{Stage 2}} &
  \multicolumn{1}{c|}{\cellcolor[HTML]{DAE8FC}0.75} &
  \multicolumn{1}{c|}{\cellcolor[HTML]{DAE8FC}0.83} &
  \multicolumn{1}{c|}{\cellcolor[HTML]{DAE8FC}0.88} &
  0.79 &
  \multicolumn{1}{c|}{\cellcolor[HTML]{9AFF99}0.61} &
  \multicolumn{1}{c|}{\cellcolor[HTML]{9AFF99}0.62} &
  \multicolumn{1}{c|}{\cellcolor[HTML]{9AFF99}-} &
  0.61 \\ \hline
\multicolumn{1}{|c|}{\cellcolor[HTML]{FFFFC7}\textbf{Stage 3}} &
  \multicolumn{1}{c|}{\cellcolor[HTML]{DAE8FC}0.86} &
  \multicolumn{1}{c|}{\cellcolor[HTML]{DAE8FC}0.86} &
  \multicolumn{1}{c|}{\cellcolor[HTML]{DAE8FC}0.95} &
  0.86 &
  \multicolumn{1}{c|}{\cellcolor[HTML]{9AFF99}0.62} &
  \multicolumn{1}{c|}{\cellcolor[HTML]{9AFF99}0.62} &
  \multicolumn{1}{c|}{\cellcolor[HTML]{9AFF99}-} &
  0.62 \\ \hline
\end{tabular}%
}
}
\caption{Comparison of McROP DPFRr-CLAHE InceptionResv2 vs Ding et al., Hybrid}
\label{table:TabledatasetsC2Stages}
\end{table}

\begin{table}[!ht]
\centering
\scalebox{0.5}{
\resizebox{\textwidth}{!}{%
\begin{tabular}{
>{\columncolor[HTML]{FFFFC7}}c |
>{\columncolor[HTML]{DAE8FC}}c 
>{\columncolor[HTML]{DAE8FC}}c 
>{\columncolor[HTML]{DAE8FC}}c 
>{\columncolor[HTML]{DAE8FC}}c |
>{\columncolor[HTML]{9AFF99}}l 
>{\columncolor[HTML]{9AFF99}}l 
>{\columncolor[HTML]{9AFF99}}l 
>{\columncolor[HTML]{9AFF99}}l |}
\cline{2-9}
\multicolumn{1}{l|}{\cellcolor[HTML]{FFFFFF}} &
  \multicolumn{4}{c|}{\cellcolor[HTML]{DAE8FC}\textbf{McROP}} &
  \multicolumn{4}{c|}{\cellcolor[HTML]{9AFF99}\textbf{Ding et al., Classifier Only}} \\ \hline
\multicolumn{1}{|c|}{\cellcolor[HTML]{FFFE65}\textbf{Stages}} &
  \multicolumn{1}{c|}{\cellcolor[HTML]{C0C0C0}\textbf{Precision}} &
  \multicolumn{1}{c|}{\cellcolor[HTML]{C0C0C0}\textbf{Recall}} &
  \multicolumn{1}{c|}{\cellcolor[HTML]{C0C0C0}\textbf{Specificity}} &
  \cellcolor[HTML]{C0C0C0}\textbf{F1} &
  \multicolumn{1}{c|}{\cellcolor[HTML]{C0C0C0}\textbf{Precision}} &
  \multicolumn{1}{c|}{\cellcolor[HTML]{C0C0C0}\textbf{Recall}} &
  \multicolumn{1}{c|}{\cellcolor[HTML]{C0C0C0}\textbf{Specificity}} &
  \multicolumn{1}{c|}{\cellcolor[HTML]{C0C0C0}\textbf{F1}} \\ \hline
\multicolumn{1}{|c|}{\cellcolor[HTML]{FFFFC7}\textbf{Stage 0}} &
  \multicolumn{1}{c|}{\cellcolor[HTML]{DAE8FC}0.67} &
  \multicolumn{1}{c|}{\cellcolor[HTML]{DAE8FC}0.63} &
  \multicolumn{1}{c|}{\cellcolor[HTML]{DAE8FC}0.88} &
  0.65 &
  \multicolumn{1}{c|}{\cellcolor[HTML]{9AFF99}-} &
  \multicolumn{1}{c|}{\cellcolor[HTML]{9AFF99}-} &
  \multicolumn{1}{c|}{\cellcolor[HTML]{9AFF99}-} &
  \multicolumn{1}{c|}{\cellcolor[HTML]{9AFF99}-} \\ \hline
\multicolumn{1}{|c|}{\cellcolor[HTML]{FFFFC7}\textbf{Stage 1}} &
  \multicolumn{1}{c|}{\cellcolor[HTML]{DAE8FC}0.67} &
  \multicolumn{1}{c|}{\cellcolor[HTML]{DAE8FC}0.6} &
  \multicolumn{1}{c|}{\cellcolor[HTML]{DAE8FC}0.94} &
  0.63 &
  \multicolumn{1}{l|}{\cellcolor[HTML]{9AFF99}0.98} &
  \multicolumn{1}{l|}{\cellcolor[HTML]{9AFF99}0.36} &
  \multicolumn{1}{l|}{\cellcolor[HTML]{9AFF99}-} &
  0.53 \\ \hline
\multicolumn{1}{|c|}{\cellcolor[HTML]{FFFFC7}\textbf{Stage 2}} &
  \multicolumn{1}{c|}{\cellcolor[HTML]{DAE8FC}0.75} &
  \multicolumn{1}{c|}{\cellcolor[HTML]{DAE8FC}0.83} &
  \multicolumn{1}{c|}{\cellcolor[HTML]{DAE8FC}0.88} &
  0.79 &
  \multicolumn{1}{l|}{\cellcolor[HTML]{9AFF99}0.45} &
  \multicolumn{1}{l|}{\cellcolor[HTML]{9AFF99}0.86} &
  \multicolumn{1}{l|}{\cellcolor[HTML]{9AFF99}-} &
  0.59 \\ \hline
\multicolumn{1}{|c|}{\cellcolor[HTML]{FFFFC7}\textbf{Stage 3}} &
  \multicolumn{1}{c|}{\cellcolor[HTML]{DAE8FC}0.86} &
  \multicolumn{1}{c|}{\cellcolor[HTML]{DAE8FC}0.86} &
  \multicolumn{1}{c|}{\cellcolor[HTML]{DAE8FC}0.95} &
  0.86 &
  \multicolumn{1}{l|}{\cellcolor[HTML]{9AFF99}0.59} &
  \multicolumn{1}{l|}{\cellcolor[HTML]{9AFF99}0.36} &
  \multicolumn{1}{l|}{\cellcolor[HTML]{9AFF99}-} &
  0.45 \\ \hline
\end{tabular}%
}
}
\caption{Comparison of McROP DPFRr-CLAHE InceptionResv2 vs Ding et al., Classifier Only}
\label{table:TabledatasetsC3Stages}
\end{table}

\begin{table}[!ht]
\centering
\scalebox{0.5}{
\resizebox{\textwidth}{!}{%
\begin{tabular}{c|
>{\columncolor[HTML]{DAE8FC}}c 
>{\columncolor[HTML]{DAE8FC}}c 
>{\columncolor[HTML]{DAE8FC}}c 
>{\columncolor[HTML]{DAE8FC}}c |
>{\columncolor[HTML]{9AFF99}}c 
>{\columncolor[HTML]{9AFF99}}c 
>{\columncolor[HTML]{9AFF99}}c 
>{\columncolor[HTML]{9AFF99}}c |}
\cline{2-9}
\multicolumn{1}{l|}{} &
  \multicolumn{4}{c|}{\cellcolor[HTML]{DAE8FC}\textbf{McROP}} &
  \multicolumn{4}{c|}{\cellcolor[HTML]{9AFF99}\textbf{Tong et al.,}} \\ \hline
\multicolumn{1}{|l|}{\cellcolor[HTML]{FFFC9E}\textbf{Stages}} &
  \multicolumn{1}{c|}{\cellcolor[HTML]{C0C0C0}\textbf{Precision}} &
  \multicolumn{1}{c|}{\cellcolor[HTML]{C0C0C0}\textbf{Recall}} &
  \multicolumn{1}{c|}{\cellcolor[HTML]{C0C0C0}\textbf{Specificity}} &
  \cellcolor[HTML]{C0C0C0}\textbf{F1} &
  \multicolumn{1}{c|}{\cellcolor[HTML]{C0C0C0}\textbf{Precision}} &
  \multicolumn{1}{c|}{\cellcolor[HTML]{C0C0C0}\textbf{Recall}} &
  \multicolumn{1}{c|}{\cellcolor[HTML]{C0C0C0}\textbf{Specificity}} &
  \cellcolor[HTML]{C0C0C0}\textbf{F1-Score} \\ \hline
\multicolumn{1}{|c|}{\cellcolor[HTML]{FFFFC7}\textbf{Stage 0}} &
  \multicolumn{1}{c|}{\cellcolor[HTML]{DAE8FC}0.67} &
  \multicolumn{1}{c|}{\cellcolor[HTML]{DAE8FC}0.63} &
  \multicolumn{1}{c|}{\cellcolor[HTML]{DAE8FC}0.88} &
  0.65 &
  \multicolumn{1}{c|}{\cellcolor[HTML]{9AFF99}-} &
  \multicolumn{1}{c|}{\cellcolor[HTML]{9AFF99}-} &
  \multicolumn{1}{c|}{\cellcolor[HTML]{9AFF99}-} &
  - \\ \hline
\multicolumn{1}{|c|}{\cellcolor[HTML]{FFFFC7}\textbf{Stage 1}} &
  \multicolumn{1}{c|}{\cellcolor[HTML]{DAE8FC}0.67} &
  \multicolumn{1}{c|}{\cellcolor[HTML]{DAE8FC}0.6} &
  \multicolumn{1}{c|}{\cellcolor[HTML]{DAE8FC}0.94} &
  0.63 &
  \multicolumn{1}{c|}{\cellcolor[HTML]{9AFF99}0.86} &
  \multicolumn{1}{c|}{\cellcolor[HTML]{9AFF99}0.77} &
  \multicolumn{1}{c|}{\cellcolor[HTML]{9AFF99}0.88} &
  0.81 \\ \hline
\multicolumn{1}{|c|}{\cellcolor[HTML]{FFFFC7}\textbf{Stage 2}} &
  \multicolumn{1}{c|}{\cellcolor[HTML]{DAE8FC}0.75} &
  \multicolumn{1}{c|}{\cellcolor[HTML]{DAE8FC}0.83} &
  \multicolumn{1}{c|}{\cellcolor[HTML]{DAE8FC}0.88} &
  0.79 &
  \multicolumn{1}{c|}{\cellcolor[HTML]{9AFF99}0.49} &
  \multicolumn{1}{c|}{\cellcolor[HTML]{9AFF99}0.55} &
  \multicolumn{1}{c|}{\cellcolor[HTML]{9AFF99}0.97} &
  0.52 \\ \hline
\multicolumn{1}{|c|}{\cellcolor[HTML]{FFFFC7}\textbf{Stage 3}} &
  \multicolumn{1}{c|}{\cellcolor[HTML]{DAE8FC}0.86} &
  \multicolumn{1}{c|}{\cellcolor[HTML]{DAE8FC}0.86} &
  \multicolumn{1}{c|}{\cellcolor[HTML]{DAE8FC}0.95} &
  0.86 &
  \multicolumn{1}{c|}{\cellcolor[HTML]{9AFF99}0.55} &
  \multicolumn{1}{c|}{\cellcolor[HTML]{9AFF99}0.47} &
  \multicolumn{1}{c|}{\cellcolor[HTML]{9AFF99}0.98} &
  0.51 \\ \hline
\end{tabular}%
}
}
\caption{Comparison of McROP DPFRr-CLAHE InceptionResv2 vs Tong et al.,}
\label{table:TabledatasetsC4Stages}
\end{table}

 With Stage only based comparison, our results in Stage 2 and Stage 3 showed significantly better 
 outcomes than Ding et al., \cite{ding2020retinopathy} Tong et al., \cite{tong2020automated}. Tables~\ref{table:TabledatasetsC2Stages},~\ref{table:TabledatasetsC3Stages}, and~\ref{table:TabledatasetsC4Stages} illustrate the comparisons between our results with theirs. Ding et al., \cite{ding2020retinopathy} used R-CNN hybrid classifier and a pure classifier. Using an R-CNN hybrid, they achieved the following sensitivity/precision: 62\%/61\% for Stage 2, and 62\%/62\% in Stage 3 in terms of sensitivity/precision .  Pure classifier results are 36\%/98\%, 86\%/45\%, and 36\%/59\%. In comparison, we performed better in classifier only in all stages while in the R-CNN hybrid, our results were better for Stage 2 and Stage 3. Ding's Stage 1 results were slightly better at 77\%/78\% using R-CNN hybrid. Further, we obtained better results using smaller 224x224 images whereas Ding et al., \cite{ding2020retinopathy} used larger 299x299 images. Similarly Tong et al., \cite{tong2020automated} used 224x224 like us and noted challenges with Stage 0 and Stage 1 similar to our study. They did not state if their Stage 1 data was inclusive of faint demarcation line as they did not show Stage 0 either or it was only Stage 1 where the demarcation line is visible. In our case, we did not distinguish between the intensity of the line pertaining to Stage 1. For Stage 2 and 3, we showed significant improvement in comparison. DPFR-CLAHE/InceptionResv2 CNN combination achieved in terms of sensitivity/specificity 83\%/88\% vs 55\%/97\%for Stage 2, 86\%/96\% vs 47\%/98\% for Stage 3.

As noted before Zones, classification was a proof of concept and this approach also offered a promising improvement which required further work to be able determine the geometric aspect zones. Further breakdown analysis of DPFRr/ResNet50 results identified Zone I/II with very high precision/sensitivity and overall accuracy. In terms of Zone I detection, it showed precision and sensitivity of 67\%/100\% (accuracy of 93\%), and Zone II with 100\%/85\% (accuracy of 85.37\%) respectively. Zhao \cite{zhao2019deep} in comparison only obtained 91\% for Zone I. This is, as far as we are aware, the only paper where Zone I and II have been shown to be classified successfully using deep learning and using restoration based methods. Zone III was challenged due to lack of sufficient data. Overall, there was improvements noted for PCAr when combined with CLAHE. Similar, same results were noted for DPFRr-CLAHE as well.

\subsection{Limitations}

This study had several limitations. We noted the challenges in obtaining large Retcam ROP datasets. Our CNNs were limited by the data we used for training including unavailable data for Stage 4 and 5 as well as Zones. Augmentation was used to generate further training data. We could not obtain independent ROP Retcam data to be used as a testing dataset, thereby limiting our ability to including testing. Our system at present allowed the classification of Plus, Stages 0-3 and to a limited degree Zones. Our future work would include the ability to firstly qualify the incoming data in terms of quality and allow it to predict all three aspects at the same time. 

\subsection{Miscellaneous findings}
As part of this series of experiments, we also investigated the possibility of training Stages classifiers with Stages training Retcam ROP dataset pre-processed using one method with corresponding Stages Retcam fundus validation dataset pre-processed using an alternate method. For Stages, training was performed with Grayscale but validation with PCAr-CLAHE.  The results are noted in Table ~\ref{table:TabledatasetsMixedStages} and its resulting confusion matrix in Table ~\ref{table:TabledatasetsMixedStagesCM}.

\begin{table}[!ht]
\centering
\resizebox{\columnwidth}{!}{%
\begin{tabular}{llllllll}
\cline{2-8}
\multicolumn{1}{l|}{} & \multicolumn{7}{c|}{\cellcolor[HTML]{ECF4FF}\textbf{InceptionResv2-Stages}} \\ \hline
\rowcolor[HTML]{EFEFEF} 
\multicolumn{1}{|c|}{\cellcolor[HTML]{EFEFEF}\textbf{Methods}} &
  \multicolumn{1}{l|}{\cellcolor[HTML]{EFEFEF}\textbf{Sensitivity}} &
  \multicolumn{1}{l|}{\cellcolor[HTML]{EFEFEF}\textbf{Specificity}} &
  \multicolumn{1}{l|}{\cellcolor[HTML]{EFEFEF}\textbf{Precision}} &
  \multicolumn{1}{l|}{\cellcolor[HTML]{EFEFEF}\textbf{F1}} &
  \multicolumn{1}{l|}{\cellcolor[HTML]{EFEFEF}\textbf{Kappa}} &
  \multicolumn{1}{l|}{\cellcolor[HTML]{EFEFEF}\textbf{AUC}} &
  \multicolumn{1}{l|}{\cellcolor[HTML]{EFEFEF}\textbf{Accuracy}} \\ \hline
\rowcolor[HTML]{ECF4FF} 
\multicolumn{1}{|l|}{\cellcolor[HTML]{ECF4FF}Gray/PCAr\_Clahe} &
  \multicolumn{1}{l|}{\cellcolor[HTML]{ECF4FF}0.1724} &
  \multicolumn{1}{l|}{\cellcolor[HTML]{ECF4FF}0.8448} &
  \multicolumn{1}{l|}{\cellcolor[HTML]{ECF4FF}0.1818} &
  \multicolumn{1}{l|}{\cellcolor[HTML]{ECF4FF}0.7327} &
  \multicolumn{1}{l|}{\cellcolor[HTML]{ECF4FF}0.176} &
  \multicolumn{1}{l|}{\cellcolor[HTML]{ECF4FF}0.5025} &
  \multicolumn{1}{l|}{\cellcolor[HTML]{ECF4FF}0.7327} \\ \hline
                      &           &          &          &          &          &          &         
\end{tabular}%
}
\caption{Mixed pre-processor use with McROP InceptionResv2}
\label{table:TabledatasetsMixedStages}
\end{table}

\begin{table}[!ht]
\centering
\resizebox{\columnwidth}{!}{%
\begin{tabular}{|
>{\columncolor[HTML]{ECF4FF}}l llll|}
\hline
\multicolumn{5}{|c|}{\cellcolor[HTML]{ECF4FF}\textbf{Confusion   Matrix}} \\ \hline
\multicolumn{1}{|l|}{\cellcolor[HTML]{ECF4FF}\textbf{Stages}} &
  \multicolumn{1}{l|}{\cellcolor[HTML]{ECF4FF}\textbf{Stage 0}} &
  \multicolumn{1}{l|}{\cellcolor[HTML]{ECF4FF}\textbf{Stage 1}} &
  \multicolumn{1}{l|}{\cellcolor[HTML]{ECF4FF}\textbf{Stage 2}} &
  \cellcolor[HTML]{ECF4FF}\textbf{Stage 3} \\ \hline
\multicolumn{1}{|l|}{\cellcolor[HTML]{ECF4FF}\textbf{Stage 0}} &
  \multicolumn{1}{l|}{\cellcolor[HTML]{ECF4FF}1} &
  \multicolumn{1}{l|}{13} &
  \multicolumn{1}{l|}{0} &
  1 \\ \hline
\multicolumn{1}{|l|}{\cellcolor[HTML]{ECF4FF}\textbf{Stage 1}} &
  \multicolumn{1}{l|}{1} &
  \multicolumn{1}{l|}{\cellcolor[HTML]{ECF4FF}7} &
  \multicolumn{1}{l|}{0}&
  1 \\ \hline
\multicolumn{1}{|l|}{\cellcolor[HTML]{ECF4FF}\textbf{Stage 2}} &
  \multicolumn{1}{l|}{1} &
  \multicolumn{1}{l|}{18} &
  \multicolumn{1}{l|}{\cellcolor[HTML]{ECF4FF}1} &
  2 \\ \hline
\multicolumn{1}{|l|}{\cellcolor[HTML]{ECF4FF}\textbf{Stage 3}} &
  \multicolumn{1}{l|}{0} &
  \multicolumn{1}{l|}{11} &
  \multicolumn{1}{l|}{0} &
  \cellcolor[HTML]{ECF4FF}3 \\ \hline
\end{tabular}%
}
\caption{Confusion Matrix for mixed pre-processor usage with McROP InceptionResv2}
\label{table:TabledatasetsMixedStagesCM}
\end{table}

The results were very poor and therefore leading to the recommendation that the same pre-processing method be used for both  Training and Validation datasets. As the resulting images were different for each pre-processing method, each of these methods can be used as additional methods for Retcam fundus image augmentation. In particular, Pixel Colour Amplification Illumination for ROP (PCAr) correction method we contributed here could be further extended to generate a set of images from the eight methods A-D and W-Z for only those where visible features are present while negating others. A wider spectrum of these now readily available methods could greatly contribute to meeting challenges of augmentation faced only within the ROP CNN arena but other fundus specific pathology areas.

\section{Conclusion}
\label{sec:ConclusionFutureWorks}

The results demonstrate that two improved novel hybrid methods for image pre-processing play a critical role in improving the accuracy of deep learning CNNs created for ROP detection for Plus, Stage, and Zone.  We also described a new method of seamlessly eroding blood vessels in a Retcam fundus image for Stages diagnosis which can be used in a clinical or other pathology identification. At the time of writing, this paper is the first to study ROP Plus, Stages and Zones classification based on the application of various image pre-processing and deep learning based on transfer learning. This paper also highlighted the challenges of re-sizing and recommended the use of lanzcos method. The image pre-processing methods noted can also be used for augmentation vs the traditional methods. This study was severely hampered by the small number of images we were able to use. In spite of this, the final results are competitive with the latest published results for ROP classification using CNNs. Our results results reflect a new way of pre-processing Retcam ROP images. These pre-processing methods can also be incorporate into telehealth or expanded for use with smartphone captured images.

To further improve detection and accuracy, we propose a number of changes specifically to create a hybrid approach using \textbf{PCAr} and \textbf{DPFRr}. Also, we will use either Mask R-CNN or Yolov to segment fundus image to isolate the non-vascular region where a demarcation line may be present. This, with further segmentation of Optic Disc (OD) may then be used for Zone III detection. Our classifiers for Plus, Stages, Zones may also be combined as a multi-instance classifier using ensemble techniques. These novel methods should also be reviewed if they may also qualify prior disqualified images using traditional image pre-processing methods, that were removed for quality reasons. If these disqualified images can be re-include them, we can increase the training/validation datasets significantly. Transfer learning using Imagenet can also enhanced by adding training from public adult fundus labelled datasets such as EYEPACS. Lastly, there is a need for Data Quality Filter process for Plus, Stages and Zones to automatically assess images to a prescribed agreed upon standard across the ROP community, including an open ROP standardized dataset with labelled ground truth. We hope to contribute to these aspects in future work.

\section{Acknowledgements}
\label{sec:Acknowledgements}

Sajid Rahim recognizes his co-supervisors, Dr Kourosh Sabri, Pediatric Ophthalmologist/Professor of Medicine at McMaster Children’s hospital who brought in specialized insights into this pediatric challenge and Dr Anna Ells, University of Calgary for supporting this study by making available ROP dataset. He also acknowledges and thanks Alex Gaudio[32] for detailed blackboard deep discussions in fundus image processing. Dr David Wong, Head of Ophthalmology/Professor of Medicine (University of Toronto) at St. Michael’s Hospital, Toronto who triggered the idea to study deep learning applicability for retinal diseases.

\bibliographystyle{ACM-Reference-Format}
\bibliography{bib}

\end{document}